\def\cleanversion{}
\documentclass[
  reprint,
  amsmath,amssymb,
  aps,physrev,
  superscriptaddress,
  footinbib,
  floatfix,
]{revtex4-2}

\usepackage{graphicx}
\graphicspath{{figures/}}
\usepackage{bm}
\usepackage[svgnames]{xcolor}
\colorlet{BLUE}{blue}
\colorlet{BLACK}{black}
\usepackage{physics}
\usepackage{booktabs}
\usepackage[caption=false]{subfig}
\usepackage{capt-of}
\usepackage{placeins}
\usepackage[normalem]{ulem}
\usepackage[colorlinks=true,allcolors=black,citecolor=DarkBlue,
linkcolor=DarkBlue,urlcolor=DarkBlue]{hyperref}

\newif\ifmarkedversion
\ifdefined\cleanversion
\markedversionfalse
\else
\markedversiontrue
\fi
\newcommand{\revision}[1]{\ifmarkedversion\textcolor{blue}{#1}\else\textcolor{black}{#1}\fi}
\newcommand{\revcolor}{\ifmarkedversion\color{blue}\else\color{black}\fi}
\newcommand{\deletion}[1]{\ifmarkedversion\textcolor{red}{\sout{#1}}\fi}

\newcommand{\revheading}[1]{\texorpdfstring{\revision{#1}}{#1}}
\newcommand{\figurefile}[2][\linewidth]{%
  \includegraphics[width=#1]{#2}%
}

\begin{document}

\title{\textbf{\revision{Matrix-product-state-assisted variational Gibbs-state preparation}}}

\author{Rui-Hao Li}
\email{lir9@ccf.org}
\affiliation{Computational Life Sciences, Cleveland Clinic Research, Cleveland, OH 44195, USA}

\author{Semeon Valgushev}
\affiliation{Department of Physics, National Tsing Hua University, Hsinchu 30013, Taiwan}
\affiliation{Department of Physics and Astronomy, Iowa State University, Ames, IA, 50011, USA}

\author{Khadijeh Najafi}
\affiliation{IBM Quantum, IBM T.J. Watson Research Center, Yorktown Heights, NY 10598 USA}
\affiliation{MIT-IBM Watson AI Lab, Cambridge, MA 02142, USA}

\begin{abstract}
  \revision{Evaluating the von Neumann entropy is a central challenge in variational Gibbs-state preparation.
    We introduce and benchmark matrix-product-state (MPS) assistance for classically evaluating and optimizing the purification circuits without full-statevector storage, examining how circuit architecture affects both thermal-state accuracy and the computational resources required.
    Comparisons on four- and six-spin transverse-field Ising and XXZ chains reveal complementary strengths: a thermofield-double ansatz performs better at high temperature, whereas a contiguous hardware-efficient ansatz (HEA) performs better on cooling.
    We identify a layer-dependent entropy ceiling in the contiguous HEA and compare it with an interleaved architecture whose entropy capacity is set solely by the ancilla qubit count.
    MPS-based benchmarks on $10\times1$, $20\times1$, $3\times3$, and $4\times4$ transverse-field Ising systems show that the tested interleaved circuits generally lower the free-energy errors and improve thermal observables.
    However, greater entropy capacity and lower free energy do not guarantee improvement in every observable.
    The architectures also differ in classical cost: contiguous registers permit entropy evaluation at a single MPS bond, whereas the interleaved arrangement requires a dense entropy calculation that scales exponentially with the number of ancillas.
  These benchmarks connect the accuracy of MPS-assisted Gibbs-state preparation to circuit capacity, optimization, and the tractability of energy and entropy evaluation.}
\end{abstract}

\maketitle

\section{Introduction}

The preparation of quantum Gibbs or thermal states is a cornerstone of quantum computation, enabling a wide array of applications in quantum technologies.
Thermal states, which represent systems in thermal equilibrium at finite temperatures, are crucial for simulating complex quantum systems, such as those in condensed matter physics, to investigate material properties at non-zero temperatures~\cite{georgescuQuantumSimulation2014,alhambraQuantumManyBodySystems2023}.
Beyond physical simulations, Gibbs states are vital in quantum machine learning, particularly for training quantum Boltzmann machines~\cite{aminQuantumBoltzmannMachine2018,zoufalVariationalQuantumBoltzmann2021}, and in quantum optimization, where sampling from well-prepared thermal states facilitates solving combinatorial optimization problems and semi-definite programming~\cite{kirkpatrickOptimizationSimulatedAnnealing1983,brandaoQuantumSpeedUpsSolving2017}.

It is known that preparing a ground state of generic quantum many body systems is QMA hard~\cite{kempeComplexityLocalHamiltonian2006,oliveiraComplexityQuantumSpin2008}.
It is generally believed that low-temperature Gibbs state preparation can be as computationally demanding as ground state preparation, necessitating sophisticated algorithms to manage the exponential complexity~\cite{chenQuantumThermalState2023}.
These challenges underscore the need for efficient and noise-resilient quantum algorithms to enable scalable thermal state preparation on quantum computers.
To address this, algorithms based on quantum phase estimation~\cite{poulinSamplingThermalQuantum2009}, linear combinations of unitaries~\cite{chowdhuryQuantumAlgorithmsGibbs2016}, quantum imaginary time evolution (QITE)~\cite{mottaDeterminingEigenstatesThermal2020}, and Lindbladian dynamics~\cite{wocjanSzegedyWalkUnitaries2023,rallThermalStatePreparation2023,chenFastThermalizationEigenstate2023,linDissipativePreparationManyBody2025} have been proposed, sometimes offering provable performance guarantees.
However, many implementations of these methods involve complex subroutines that require deep circuits and extensive quantum resources, limiting their practicality for near-term quantum devices.

Variational quantum algorithms (VQAs) provide a promising route for preparing quantum Gibbs states on noisy intermediate-scale quantum (NISQ) devices through shallow circuits and hybrid quantum-classical optimization.
As discussed in detail in Sec.~\revision{\ref{sec:framework}}, variational Gibbs state preparation typically involves parameterizing a trial mixed state using a quantum circuit and optimizing the circuit parameters to minimize the Helmholtz free energy.
However, evaluating the free energy remains challenging since the evaluation of the von Neumann entropy typically requires resource-intensive tomography or stochastic reconstruction~\cite{teoQuantumStateReconstructionMaximizing2011,banaszekFocusQuantumTomography2013,guptaMaximalEntropyApproach2021}.
To overcome this obstacle, several approaches have been proposed.
Many of the existing methods employ the thermofield double (TFD) state purification~\cite{wuVariationalThermalQuantum2019,zhuGenerationThermofieldDouble2020}, in which two entangled replicas of the system encode thermal correlations as a pure state.
Based on this idea, one strategy is to use alternative cost functions that avoid direct entropy estimation, such as the engineered cost function extrapolated from small-size systems~\cite{sagastizabalVariationalPreparationFinitetemperature2021}, one involving purity measurements~\cite{warrenAdaptiveVariationalAlgorithms2022}, and one based on truncated Taylor-series expansions of the free energy operator~\cite{wangVariationalQuantumGibbs2021}.
Another strategy employs a smart modular construction of the ansatz that allows for estimation of the von Neumann entropy through classical post-processing of measurement outcomes~\cite{consiglioVariationalGibbsState2024,sambasivamTEPIDADAPTAdaptiveVariational2025}.
Furthermore, approaches such as variational QITE~\cite{zoufalVariationalQuantumBoltzmann2021,wangCriticalBehaviorIsing2023} and dissipative algorithms~\cite{ilinDissipativeVariationalQuantum2025} have been explored as an alternative to direct free energy minimization.

NISQ-compatible variational approaches for quantum Gibbs-state preparation have been demonstrated on quantum hardware using small lattice models~\cite{sagastizabalVariationalPreparationFinitetemperature2021,francisManybodyThermodynamicsQuantum2021,consiglioVariationalGibbsState2024}.
A complementary adiabatic approach to thermal state preparation for a $5\times4$ Ising model was recently demonstrated on a trapped-ion quantum processor, marking a notable advancement in system size~\cite{granetAdiabaticPreparationThermal2025}.
Nonetheless, efficiently preparing Gibbs states for larger systems continues to be a challenge on current quantum devices.
\revision{Scaling variational preparation requires affordable free-energy evaluation as well as implementable circuits, which motivates classical assistance for the energy and entropy calculations.}

To this end, tensor-network techniques, particularly matrix product states (MPSs) and matrix product operators (MPOs), can provide compact classical representations of states with tractable bond dimensions.
Tensor-network methods have been used to benchmark the expressiveness of variational circuit architectures~\cite{haghshenasVariationalPowerQuantum2022} and to construct reduced-depth circuits for quantum dynamics and MPS preparation~\cite{zhangScalableQuantumDynamics2024,robertsonApproximateQuantumCompiling2025,jaderbergVariationalPreparationNormal2025}.
{\revcolor
  An efficient classical MPS description does not by itself provide a shallow quantum preparation circuit, motivating studies of MPS preparation and verification~\cite{malzPreparationMatrixProduct2024,cruzPreparationVerificationTensor2022,zhouSymmetryprotectedDissipativePreparation2021}.
  Here, we use MPS assistance to classically evaluate and optimize purification circuits, with the aim of assessing how their architecture constrains thermal accuracy and preparation resources.
  In particular, MPS assistance allows for the evaluation of the entropy and energy of the trial state without storing full statevectors or expensive quantum measurements, potentially extending accessible system sizes when tensor-network costs remain tractable.

  However, it is important to note that affordable evaluation of the free energy alone does not ensure accurate preparation: the accuracy of the resulting thermal state still depends on the chosen circuit and its optimization.
  For example, at high temperatures, a purification must generate sufficient entanglement between the system and ancillas to reproduce the large thermal entropy, which can be challenging for shallow circuits.
  At lower temperatures, it must capture the populations and correlations of the relevant low-energy states well enough to reproduce the target observables.
  These requirements may not be equally well accommodated by a given ansatz.
  Indeed, simulation and hardware experiments have shown that the fidelity of the prepared state with respect to the target Gibbs state can be nonmonotonic with temperature~\cite{consiglioVariationalGibbsState2024,robertsonVariationalGibbsState2026}.
  Moreover, even when the goal is to reproduce local observables rather than achieve high global fidelity, agreement with the thermal reference for one observable does not guarantee comparable accuracy for others.
}

\revision{In this work, we benchmark MPS-assisted variational Gibbs-state preparation by examining how circuit architecture, entropy capacity, and optimization affect thermal-state accuracy, and how these choices determine the classical computational resources required.}
In Sec.~\revision{\ref{sec:framework}}, we detail the MPS-assisted variational algorithm and discuss two popular ansatzes for Gibbs state preparation in the literature, a thermofield-double ansatz (TFDA) and a contiguous hardware-efficient ansatz (HEA).
We compare their performances through small-scale numerical simulations of the transverse-field Ising model (TFIM) and XXZ model.
{\revcolor
  Sec.~\ref{sec:architectures} introduces an interleaved arrangement of the HEA to address the contiguous circuit's entropy bottleneck and relates both architectures to their entropy bounds and classical costs.
  In Sec.~\ref{sec:shared-benchmarks}, we benchmark both HEAs using MPS simulations on the same $10\times1$, $20\times1$, $3\times3$, and $4\times4$ TFIM systems.
  We assess the global objective together with local and non-local observables.
  In these finite-size benchmarks, the optimization is fully classical; quantum hardware would enter only when executing selected circuits.
  Moreover, the benchmarks remain within the tractable regime of classical evaluation and do not demonstrate a quantum computational advantage for the equilibrium quantities studied.
  Finally, Sec.~\ref{sec:conclusion} summarizes the limitations and open directions.
}

\section{MPS-assisted variational algorithm}
\label{sec:framework}

\subsection{Purification and free-energy framework}

For a system with Hamiltonian $H$, thermal equilibrium at finite positive inverse temperature $\beta=1/(k_{\mathrm B}T)$ is described by the Gibbs state
\begin{equation}
  \rho_\beta=\frac{e^{-\beta H}}{Z_\beta},
  \label{eq:gibbs}
\end{equation}
where $Z_\beta=\Tr e^{-\beta H}$ is the partition function.
We set Boltzmann's constant $k_{\mathrm B}=1$.

Variationally preparing the Gibbs state hinges on estimating the Helmholtz free energy of a trial mixed state $\rho_S$, which is given by
\begin{equation}
  F_{\beta}(\rho_S)=E(\rho_S)-\beta^{-1}S(\rho_S),
  \label{eq:free-energy}
\end{equation}
where $E(\rho_S)$ is the energy and $S(\rho_S)$ is the von Neumann entropy of the trial state, which are respectively defined as
\begin{align}
  E(\rho_S)&=\Tr(\rho_S H),
  \label{eq:energy}\\
  S(\rho_S)&=-\Tr(\rho_S\ln\rho_S).
  \label{eq:entropy}
\end{align}
The target is obtained by minimizing this free energy,
\begin{equation}
  \rho_\beta=\arg\min_{\rho_S}F_{\beta}(\rho_S).
  \label{eq:gibbs-variational}
\end{equation}

\begin{figure*}[t]
  \centering
  \figurefile[0.94\textwidth]{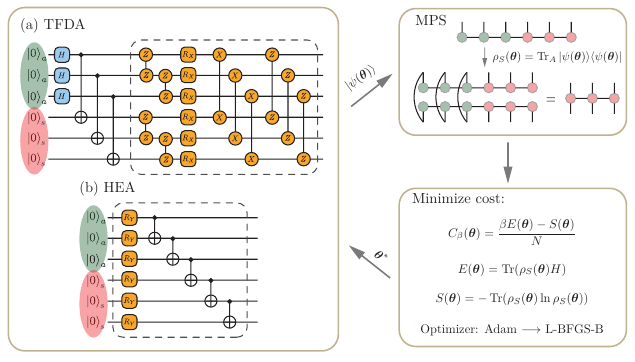}
  \caption{\normalcolor Schematic of the MPS-assisted variational Gibbs state preparation algorithm, showcasing two ansatzes benchmarked in this work: (a) the TFDA for the 1D transverse-field Ising model, and (b) the contiguous HEA.
    The green- and red-shaded ovals highlight the ancilla ($a$) and system ($s$) qubits, respectively.
    In the TFDA, the numbers of these two types of qubits are always equal, while in HEAs, the number of ancilla qubits can be adjusted independently.
    In both circuits, the orange gates represent parameterized gates, including single-qubit rotations such as $R_X$ and $R_Y$, and two-qubit entangling gates such as $R_{ZZ}$ and $R_{XX}$.
    The gray dashed boxes represent the part of circuit that can be repeated for multiple layers.
    The parameterized quantum circuit prepares a pure state $\ket{\psi(\boldsymbol{\theta})}$ on an enlarged Hilbert space containing the system and ancilla qubits.
    Its matrix-product-state (MPS) representation is built classically to evaluate the energy $E$ and von Neumann entropy $S$ of the reduced system state $\rho_S(\boldsymbol\theta)$.
    \revision{Adam followed by L-BFGS-B minimizes the normalized free energy $C_\beta(\boldsymbol\theta)$ with respect to the circuit parameters $\boldsymbol\theta$.
  The optimized circuit can subsequently be measured to provide estimates of thermal observables.}}
  \label{fig:workflow}
\end{figure*}

When the trial state $\rho_S$ is prepared by a parameterized quantum circuit, such as the ones shown in Fig.~\ref{fig:workflow}, we denote the state as $\rho_S(\boldsymbol\theta)$, where $\boldsymbol\theta$ are the variational parameters in the circuit.
While the energy term $\Tr(\rho_S H)$ can be estimated rather efficiently via quantum measurements, especially for simple spin models, von Neumann entropy estimation poses a critical bottleneck: estimation of $S(\rho_S)$ typically relies on at least partial state tomography of $\rho_S$, or various approximation schemes, which are often inefficient and/or inaccurate~\cite{wangVariationalQuantumGibbs2021,warrenAdaptiveVariationalAlgorithms2022,sambasivamTEPIDADAPTAdaptiveVariational2025}.
For near-term devices, this is further exacerbated by measurement noise and limited qubit coherence.

To potentially overcome this challenge, we propose to leverage MPS to classically approximate the state of the parameterized quantum circuit, which serves as the trial purified Gibbs state.
Generally speaking, purification embeds the mixed state $\rho_S$ into a pure state $\ket\psi$ on an enlarged Hilbert space, with
\begin{equation}
  \rho_S(\boldsymbol\theta)=\Tr_A
  \ket{\psi(\boldsymbol\theta)}\!\bra{\psi(\boldsymbol\theta)},
  \label{eq:purification}
\end{equation}
where the ancilla qubits are traced out.
There are $N$ system qubits in register $S$ and $N_a$ ancilla qubits in register $A$; the input-ket subscripts $s$ and $a$ in the figures denote these same registers.
MPS is a tensor network that introduces a low-rank approximation of quantum states, allowing for efficient representation and manipulation of them~\cite{orusPracticalIntroductionTensor2014}.
A clear advantage of MPS is that the number of parameters scales just linearly with the system size, at fixed bond dimension, in contrast to a general quantum state, which scales exponentially.
Therefore, by representing $\ket\psi$ as an MPS, we compute $E(\rho_S)$ and $S(\rho_S)$ classically, avoiding costly quantum measurements and state tomography for the entropy estimation.
Gates are propagated on the MPS tensors and energy is obtained by contraction with the Hamiltonian, without forming a full density matrix.

The von Neumann entropy is computed by first taking the Schmidt decomposition of the pure state $\ket\psi$ across the system--ancilla bipartition:
\begin{equation}
  \ket{\psi(\boldsymbol\theta)}
  =\sum_j\lambda_j(\boldsymbol\theta)\ket{u_j}_S\ket{v_j}_A.
  \label{eq:schmidt}
\end{equation}
Then when we trace out the ancilla qubits to get $\rho_S$, the resulting eigenvalues are precisely the squared Schmidt coefficients, $p_j=\lambda_j^2$. Therefore, von Neumann entropy of the system can then be computed as
\begin{equation}
  S(\boldsymbol\theta)=-\sum_j\lambda_j^2(\boldsymbol\theta)
  \ln\lambda_j^2(\boldsymbol\theta).
  \label{eq:schmidt-entropy}
\end{equation}
Here and below, $E(\boldsymbol\theta)$ and $S(\boldsymbol\theta)$ abbreviate these quantities evaluated on $\rho_S(\boldsymbol\theta)$, as in Fig.~\ref{fig:workflow}.
{\revcolor
  The entropy in Eq.~\eqref{eq:schmidt-entropy} measures entanglement between the system and ancillas, rather than between spatial regions of the system.
  For a system with a unique ground state, the thermal entropy vanishes on cooling to zero temperature, although spatial entanglement can remain nonzero.
}

As schematically shown in Fig.~\ref{fig:workflow}, which illustrates the workflow of the MPS-assisted variational algorithm, we start with a parameterized quantum circuit that prepares the pure state $\ket{\psi(\boldsymbol\theta)}$ on the enlarged Hilbert space containing the system (red-shaded) and ancilla (green-shaded) qubits.
After converting the state to its MPS representation, we contract the tensors to compute the energy and entropy of the resulting mixed state $\rho_S(\boldsymbol\theta)$ to obtain its free energy. The normalized cost used in simulations is defined in Sec.~\ref{sec:shared-benchmarks}.
The parameters $\boldsymbol\theta$ are iteratively optimized using a classical optimizer until a stopping criterion or safety limit is reached.
The selected parameters $\boldsymbol\theta^*$ can then be used to prepare the approximate Gibbs state $\tilde\rho_\beta=\rho_S(\boldsymbol\theta^*)$ on a quantum device.
\revision{All optimization in the present benchmarks is therefore classical.
Its usefulness at larger sizes depends on keeping both MPS propagation and entropy evaluation tractable.}

\subsection{Candidate ansatz families and small-system comparison}
\label{sec:small-comparison}

When it comes to variational Gibbs state preparation, the choice of ansatz plays a crucial role in the quality of the prepared state and the efficiency of the optimization process.
To benchmark the performance of different ansatzes, we first perform small-scale statevector simulations of the 1D transverse-field Ising model (TFIM) of 4 and 6 spins, whose Hamiltonian is given by
\begin{equation}
  H_{\rm I}=-J\sum_{i=1}^{N-1}Z_iZ_{i+1}
  -h\sum_{i=1}^{N}X_i,
  \label{eq:tfim}
\end{equation}
where $J$ is the coupling strength and $h$ is the transverse field.
We set $J=1$ and $h=0.5$.
We use $X_i,Y_i,Z_i$ for the Pauli operators and impose open boundary conditions.
In addition, we also perform the same simulations for the 1D XXZ Heisenberg model, whose Hamiltonian can be written as
\begin{equation}
  H_{\rm XXZ}=-J\sum_{i=1}^{N-1}
  (X_iX_{i+1}+Y_iY_{i+1}+\Delta Z_iZ_{i+1}),
  \label{eq:xxz}
\end{equation}
where $\Delta$ is the anisotropy parameter.
We set $J=1$ and $\Delta=-1.5$ with the rest of setup the same as for the TFIM.  The XXZ model is used only in this small-system comparison.

One of the earliest ansatzes proposed for this task is motivated by the quantum approximate optimization algorithm (QAOA)~\cite{farhiQuantumApproximateOptimization2014} and adiabatic evolution.
This ansatz~\cite{wuVariationalThermalQuantum2019,zhuGenerationThermofieldDouble2020}, shown in Fig.~\ref{fig:workflow}(a) and referred to as the TFDA in the following, is designed to capture the essential features of the target thermal state by leveraging the structure of the underlying Hamiltonian.
Since its inception, the TFDA has been adopted in various studies with different modifications, such as the use of an engineered cost function that is extrapolated from the target state for small-size systems~\cite{sagastizabalVariationalPreparationFinitetemperature2021} and the incorporation of the adaptive variational algorithm framework~\cite{warrenAdaptiveVariationalAlgorithms2022}.
It builds on the idea of the thermofield double (TFD) state, which is a purification of the Gibbs state on an enlarged Hilbert space that doubles the size of the system.
Tracing out either complete register in the TFD state yields the Gibbs state of the other register.

As shown in Fig.~\ref{fig:workflow}(a), in its general form, the TFDA starts by preparing the TFD state at $\beta=0$, $\ket{\mathrm{TFD}(0)}$, which is a product of system--ancilla Bell pairs.
\revision{
  With $N_a=N$, this state is
  \begin{equation}
    \ket{\mathrm{TFD}(0)}=\bigotimes_{i=1}^{N}
    \ket{\Phi^+}_{S_iA_i},
    \label{eq:tfd0}
  \end{equation}
  where
  \begin{equation}
    \ket{\Phi^+}=\frac{\ket{00}+\ket{11}}{\sqrt2}.
    \label{eq:bell-pair}
  \end{equation}
  Each pair is maximally entangled, and their product is maximally entangled across the complete $S|A$ partition.
}
It is worth noting that tracing out either register of $\ket{\mathrm{TFD}(0)}$ yields the maximally mixed state, which is exactly the Gibbs state at infinite temperature.
{\revcolor
  Starting from these Bell pairs, the variational gates change the spectrum and correlations of the reduced state.
  These gates are structured to reflect the interactions in the Hamiltonian, with separate blocks for intra-register terms and inter-register couplings.
  In our implementation, each TFDA layer has four independent angles, shared among gates of the same class within that layer.
  For a Pauli operator $Q\in\{X,Y,Z\}$ and rotation angle $\phi$, we use the following single- and two-qubit rotation gates, $R_Q(\phi)=e^{-i\phi Q/2}$ and $R_{QQ}(\phi)=e^{-i\phi(Q\otimes Q)/2}$, respectively.
  A TFIM layer first applies interactions within each register, then transverse-field rotations, and finally gates coupling paired system and ancilla qubits:
  \begin{subequations}
    \begin{align}
      U_{{\rm edge},\ell}^{\rm TFIM}
      &=\prod_{i=1}^{N-1}
      [R_{ZZ}^{S_iS_{i+1}}(\theta_{\ell0})\,
      R_{ZZ}^{A_iA_{i+1}}(\theta_{\ell0})],\label{eq:tfda-tfim-edge}\\
      U_{{\rm field},\ell}^{\rm TFIM}
      &=\prod_{i=1}^{N}
      [R_X^{S_i}(\theta_{\ell1})\,R_X^{A_i}(\theta_{\ell1})],\label{eq:tfda-tfim-field}\\
      U_{{\rm pair},\ell}
      &=\prod_{i=1}^{N}
      [R_{XX}^{S_iA_i}(\theta_{\ell2})\,
      R_{ZZ}^{S_iA_i}(\theta_{\ell3})].\label{eq:tfda-pair}
    \end{align}
  \end{subequations}
  Here we use $\ell=1,\ldots,L$ to index layers, with angles $\theta_{\ell0},\ldots,\theta_{\ell3}$.
  Superscripts on a gate identify the qubits on which it acts.
  All gates within each of these three blocks commute, so their internal order does not matter.
  The order between blocks still matters.
  With the usual right-to-left action on states, the complete layer is
  \begin{equation}
    \mathcal U^{\rm TFIM}_\ell
    =U_{{\rm pair},\ell}U_{{\rm field},\ell}^{\rm TFIM}
    U_{{\rm edge},\ell}^{\rm TFIM}.
    \label{eq:tfda-tfim-layer}
  \end{equation}
  For XXZ, the intra-register block instead contains all three interaction axes,
  \begin{equation}
    \begin{split}
      U_{{\rm edge},\ell}^{\rm XXZ}
      =\overset{\rightarrow}{\prod_{i=1}^{N-1}}[
        &R_{XX}^{S_iS_{i+1}}(\theta_{\ell0})\,
        R_{XX}^{A_iA_{i+1}}(\theta_{\ell0})\,\\
        &R_{YY}^{S_iS_{i+1}}(\theta_{\ell0})\,
        R_{YY}^{A_iA_{i+1}}(\theta_{\ell0})\,\\
        &R_{ZZ}^{S_iS_{i+1}}(\theta_{\ell1})\,
      R_{ZZ}^{A_iA_{i+1}}(\theta_{\ell1})].
    \end{split}
    \label{eq:tfda-xxz-edge}
  \end{equation}
  Here the angle $\theta_{\ell0}$ is shared among the $XX$ and $YY$ gates, while $\theta_{\ell1}$ is used for the $ZZ$ gates, reflecting the anisotropy of the XXZ Hamiltonian in Eq.~\eqref{eq:xxz}.
  The gates within each bracket commute and may be applied in any order, whereas blocks on adjacent bonds generally do not commute.
  Therefore, the order of these blocks needs to be specified.
  We use the arrow on the product to indicate that the bond blocks are applied in increasing site index in our construction.
  The intra-register block is followed by the same system--ancilla pair block as in Eq.~\eqref{eq:tfda-pair}, giving the complete XXZ layer
  \begin{equation}
    \mathcal U^{\rm XXZ}_\ell=U_{{\rm pair},\ell}U_{{\rm edge},\ell}^{\rm XXZ}.
    \label{eq:tfda-xxz-layer}
  \end{equation}
  Repeating either layer $L$ times, with separate angles for each layer, gives $4L$ independent parameters.
}

\begin{figure}[t]
  \centering
  \figurefile[\linewidth]{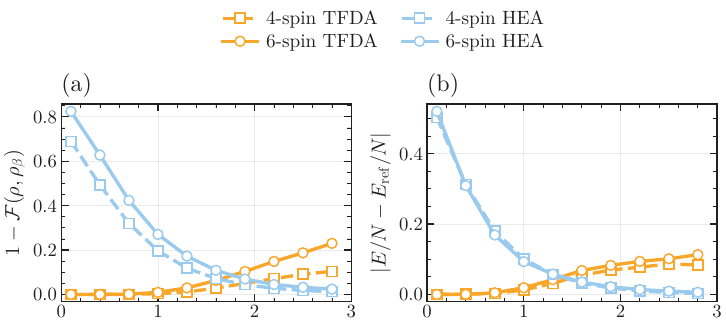}\par
  \vspace{0.25\baselineskip}
  \figurefile[\linewidth]{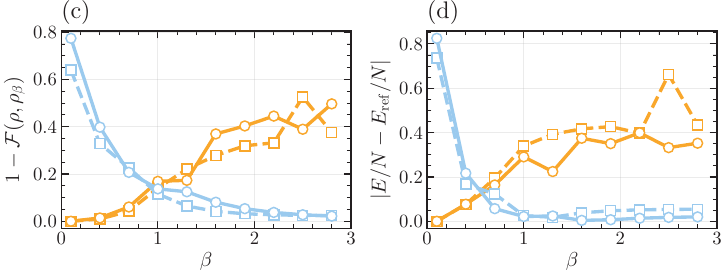}
  \caption{\normalcolor Performance comparison of the TFDA and HEA on the 1D open-boundary TFIM [panels (a) and (b)] and XXZ model [panels (c) and (d)], each of 4 and 6 spins at various inverse temperatures $\beta$.
    \revision{
      Panels (a) and (c) show the infidelity $1-\mathcal F$ relative to the exact Gibbs state, using the squared Uhlmann fidelity $\mathcal F$; panels (b) and (d) show the absolute energy-density error $|E/N-E_{\rm ref}/N|$.
    Orange and blue curves represent the TFDA and contiguous HEA, respectively, whereas squares with dashed lines denote 4 spins and circles with solid lines denote 6 spins.}
    For both ansatzes, the number of ancilla qubits used is equal to $N$, where $N$ is the number of physical spins in the system.
    The number of layers is $L = N/2$.
  Each data point is the best result from the 20 optimization runs with different initial parameters.}
  \label{fig:small-family}
\end{figure}

Another useful ansatz is the HEA~\cite{kandalaHardwareefficientVariationalQuantum2017}, shown in Fig.~\ref{fig:workflow}(b), which is designed to be compatible with the qubit connectivity of near-term quantum devices and has been studied for preparing Gibbs states of local Hamiltonians~\cite{wangVariationalQuantumGibbs2021}.
{\revcolor
  We adopt the same contiguous HEA construction as Ref.~\cite{wangVariationalQuantumGibbs2021} for our small-scale benchmarks.
  It begins from $\ket{0}^{\otimes(N+N_a)}$ in the order
  $A_1,\ldots,A_{N_a},S_1,\ldots,S_N$.
  For $n=N+N_a$, layer $\ell$ consists of a block of single-qubit $R_Y$ rotations on all qubits, followed by a staircase of CNOTs connecting qubits $q$ and $q+1$ for $q=1,\ldots,n-1$:
  \begin{subequations}
    \begin{align}
      U_R^{(\ell)}&=\bigotimes_{q=1}^{n}R_Y(\phi_{\ell q}),\label{eq:contiguous-rotations}\\
      U_C&=\overset{\rightarrow}{\prod_{q=1}^{n-1}}\mathrm{CNOT}_{q\rightarrow q+1},\label{eq:contiguous-staircase}
    \end{align}
  \end{subequations}
  where the CNOTs act in increasing $q$, and $\boldsymbol\theta$ collects the independent angles $\phi_{\ell q}$.
  The complete layer is
  \begin{equation}
    \mathcal U_\ell^{\rm cont}=U_CU_R^{(\ell)}.
    \label{eq:contiguous-layer}
  \end{equation}
  Hence, the full circuit has $nL$ parameters and $(n-1)L$ CNOT gates.

}
Since the TFDA always requires the same number of ancilla qubits as the system ones, we also adopt the same setup for the HEA, i.e., we use 4 and 6 ancilla qubits for the 4- and 6-spin systems, respectively.
Both ansatzes have $L=N/2$ layers, where $N$ is the number of spins in the system.
{\revcolor
  We evaluate ten physical inverse-temperature points, $\beta\in\{0.1,0.4,0.7,1.0,1.3,\allowbreak1.6,1.9,2.2,2.5,2.8\}$.
  The chosen inverse-temperature range spans highly mixed states near the infinite-temperature limit and colder states with substantially reduced thermal entropy.
  This allows us to assess the performance of the ansatzes across a broad spectrum of thermal states.
  We use the combined Adam and L-BFGS-B optimizers to optimize the parameters of the ansatzes, with the stopping criteria specified in Appendix~\ref{app:optimization-protocol}.
}
Moreover, for each $\beta$, we run the optimization for 20 different random initializations of the parameters to mitigate the stochasticity of the optimization process.
We then choose the best result based on the lowest \revision{independently reconstructed $C_\beta$} for each $\beta$.
\revision{Here $C_\beta$ is proportional to the free energy at fixed $N$ and $\beta$; its normalization is introduced in Sec.~\ref{sec:shared-benchmarks}.}
The results are shown in Fig.~\ref{fig:small-family}.
In the left panels [Figs.~\ref{fig:small-family}(a) and (c)], we show the infidelity of the prepared Gibbs state with respect to the exact Gibbs state, which can be computed classically for small systems, as a function of $\beta$.
The infidelity is $1-\mathcal F$, where the squared Uhlmann fidelity is
\begin{equation}
  \mathcal F(\rho,\rho_\beta)
  =\left[\Tr\sqrt{\sqrt\rho\,\rho_\beta\sqrt\rho}\right]^2,
  \label{eq:fidelity}
\end{equation}
In the right panels [Figs.~\ref{fig:small-family}(b) and (d)], we show the absolute difference between the estimated and exact thermal energies per spin, i.e., $|E/N-E_{\rm ref}/N|$, as a function of $\beta$.
Throughout the figures, ``ref'' denotes the exact thermal reference.

Comparing the two ansatzes, we find drastically different behaviors.
For the TFDA, shown as orange squares and circles in Fig.~\ref{fig:small-family}, the infidelity is relatively low near $\beta=0$ and generally increases with $\beta$.
This is also reflected in the thermal energy estimates, which show a noticeable deviation from the exact value at higher $\beta$ values.
Such behavior is expected, as the initial state of the TFDA is the maximally entangled state across the system--ancilla partition, which, after tracing out the ancilla system, yields the correct Gibbs state at $\beta=0$.
\revision{
  Indeed, setting all angles to zero in the TFDA reproduces the infinite-temperature state exactly.
  However, as $\beta$ increases, the TFD state becomes less entangled across the system--ancilla partition.
  Starting from maximally entangled Bell pairs, the circuit must therefore reduce this entanglement while reproducing the spectrum and correlations of the target Gibbs state.
  This provides an intuitive interpretation of the increasing infidelity and energy estimation error for a fixed-depth ansatz as $\beta$ increases, although the amount of entanglement to be removed does not necessarily determine the required circuit depth.
  Additional layers may improve its performance at larger $\beta$ by increasing the flexibility of the ansatz, but we do not test this depth dependence here.
  Any such improvement would come at the cost of additional gates and execution time required on near-term quantum devices.
}

On the other hand, the performance of the HEA shows the opposite trend: the infidelity is high at low $\beta$ and generally decreases as $\beta$ increases.
One relevant distinction is that the initial state of the HEA is a product state, which has no entanglement between the system and ancilla qubits.
\revision{For the contiguous architecture, a structural rank bound limits the entropy generated per layer, as discussed in Sec.~\ref{sec:architectures}.
This restriction is most severe at high temperature.}
For XXZ, similar to the TFIM, we observe that the HEA outperforms the TFDA in the low-temperature (large-$\beta$) regime, while the TFDA performs better at high temperatures (small $\beta$), within the tested range.
\revision{
  For both TFIM sizes, the first sampled point at which the HEA has lower infidelity is $\beta=1.9$, while its energy-density error becomes smaller at $\beta=1.6$.
  The corresponding first crossings for XXZ occur at $\beta=1.0$ and $0.7$, respectively.
  These comparisons concern the specified circuits, which do not share equal computational resources: the same layer number $L$ gives different gate and parameter counts in the two families.
}

In summary, the two ansatzes have their own strengths and weaknesses.
The TFDA, similar to the QAOA ansatz, is physically motivated and performs well at high temperatures in our benchmarks.
However, its inter-register gates may require routing on hardware with limited connectivity, and its initial Bell pairs require an MPS bond dimension exponential in $N$ when the two registers occupy separate blocks.
This makes the TFDA more suitable for high-temperature Gibbs-state preparation, which is less interesting for many applications as high-temperature Gibbs states are more classically tractable~\cite{bakshiHighTemperatureGibbsStates2024}.
In contrast, HEAs are designed to be compatible with near-term quantum devices and offer more flexibility.
Their ancilla number can be chosen independently of the system size, while their depth and connectivity can be adjusted to the available resources.
The same circuit structure can also be used for different local Hamiltonians without redesigning model-specific gate blocks.
These features, together with the favorable low-temperature performance in our small-system benchmarks, motivate our focus on HEAs in the remainder of this work.

\section{\revheading{HEA architectures, entropy capacity, and computational resources}}
\label{sec:architectures}

\subsection{\revheading{Contiguous and interleaved layouts}}

\begin{figure*}[t]
  \centering
  \figurefile[0.94\textwidth]{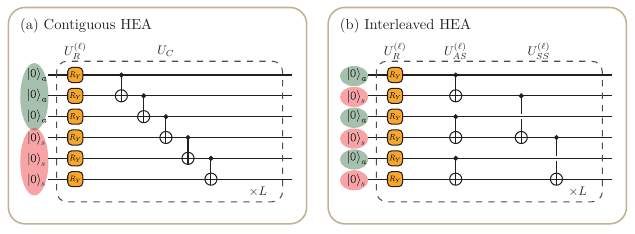}
  \caption{\normalcolor\revision{Circuit structures of the (a) contiguous and (b) interleaved HEAs, illustrated with three system qubits and three ancilla qubits.}
    The green- and red-shaded ovals highlight the ancilla ($a$) and system ($s$) qubits, respectively.
    \revision{The wires are ordered as $A_1,A_2,A_3,S_1,S_2,S_3$ in the contiguous layout and $A_1,S_1,A_2,S_2,A_3,S_3$ in the interleaved layout.}
    The gray dashed boxes represent the part of circuit that can be repeated for multiple layers.
    \revision{Each of the $L$ layers begins with parameterized single-qubit $R_Y$ rotations, shown in orange.
      In (a), the CNOT staircase $U_C$ crosses the system--ancilla partition once per layer.
  In (b), the block $U_{AS}^{(\ell)}$ couples separate ancilla--system pairs, followed by the system-only staircase $U_{SS}^{(\ell)}$, where $\ell$ labels the layer.}}
  \label{fig:hea-architectures}
\end{figure*}

{\revcolor
  Here we compare two qubit arrangements within the HEA family, which we refer to as contiguous and interleaved.
  The contiguous circuit of Eq.~\eqref{eq:contiguous-layer} places the ancillas and system qubits in separate blocks.
  As Fig.~\ref{fig:hea-architectures}(a) shows, only the CNOT between the last ancilla qubit $A_{N_a}$ and the first system qubit $S_1$ crosses the $S|A$ partition in each layer.
  Therefore, gates acting fully within either register can redistribute correlations but cannot change the Schmidt spectrum across this partition, which determines the von Neumann entropy of the reduced system state.
  This observation suggests an alternative approach: distribute the ancilla--system gates throughout the register rather than concentrating them at one boundary.

  Fig.~\ref{fig:hea-architectures}(b) illustrates the resulting interleaved layout.
  It similarly starts from $\ket{0}^{\otimes(N+N_a)}$, with $N_a\le N$.
  System sites follow chain order in one dimension and row-major order in two dimensions.
  In the latter case, we label the sites from left to right within each row before continuing with the next row in the same direction.
  For example, on a $4\times4$ lattice, the first row contains $S_1,\ldots,S_4$, the second $S_5,\ldots,S_8$, and so on.
  We distribute the ancillas approximately uniformly along this sequence by pairing ancilla $A_k$, for $k=1,\ldots,N_a$, with system site $S_{p_k}$, where
  \begin{equation}
    p_k=1+\left\lfloor\frac{(k-1)N}{N_a}\right\rfloor.
    \label{eq:partner-map}
  \end{equation}
  The MPS sequence follows $S_1,\ldots,S_N$, with each $A_k$ inserted immediately before its paired $S_{p_k}$.
  When $N_a=N$, the order is $A_1,S_1,\allowbreak A_2,S_2,\allowbreak\ldots,\allowbreak A_N,S_N$.

  Let $\mathcal Q_{\rm int}$ denote this ordered set of qubits.
  The interleaved layer consists of three disjoint blocks, as shown in Fig.~\ref{fig:hea-architectures}(b):
  \begin{subequations}
    \label{eq:interleaved-blocks}
    \begin{align}
      U_R^{(\ell)}&=\bigotimes_{q\in\mathcal Q_{\rm int}}R_Y(\phi_{\ell q}),\label{eq:interleaved-rotations}\\
      U_{AS}^{(\ell)}&=\overset{\rightarrow}{\prod_{k=1}^{N_a}}
      \mathrm{CNOT}_{A_k\rightarrow S_{p_k}},\label{eq:interleaved-as}\\
      U_{SS}^{(\ell)}&=\overset{\rightarrow}{\prod_{i=1}^{N-1}}
      \mathrm{CNOT}_{S_i\rightarrow S_{i+1}}.
      \label{eq:interleaved-ss}
    \end{align}
  \end{subequations}
  Thus an interleaved layer is
  \begin{equation}
    \mathcal U_\ell^{\rm int}=U_{SS}^{(\ell)}U_{AS}^{(\ell)}U_R^{(\ell)}.
    \label{eq:interleaved-layer}
  \end{equation}
  The $U_{AS}$ block can create entanglement through separate ancilla--system links, while $U_{SS}$ spreads correlations along the system chain.
  Interleaving can place an ancilla between consecutive system sites in the MPS ordering, giving a three-site segment $S_i,A_k,S_{i+1}$.
  The CNOT in $U_{SS}$ acts only on $S_i$ and $S_{i+1}$, with the identity acting on $A_k$.
  To represent the resulting correlations, our MPS calculation updates all three tensors and their connecting bonds exactly, without truncation or SWAP actions.
  Updating the middle tensor changes its internal bond structure but not the physical operation on the ancilla, and introduces no extra gates into the simulated circuit.
  Moreover, the final system-only block $U_{SS}^{(L)}$ leaves the eigenvalues of $\rho_S$ unchanged.
  We can therefore evaluate the final entropy immediately before this block, while retaining it when computing the energy.
  Earlier $U_{SS}$ blocks remain necessary because they influence the state entering subsequent system--ancilla gates.
}

\subsection{\revheading{Rank and entropy capacity}}
\label{sec:rank-entropy-capacity}

\begin{figure}[t]
  \centering
  \figurefile[\linewidth]{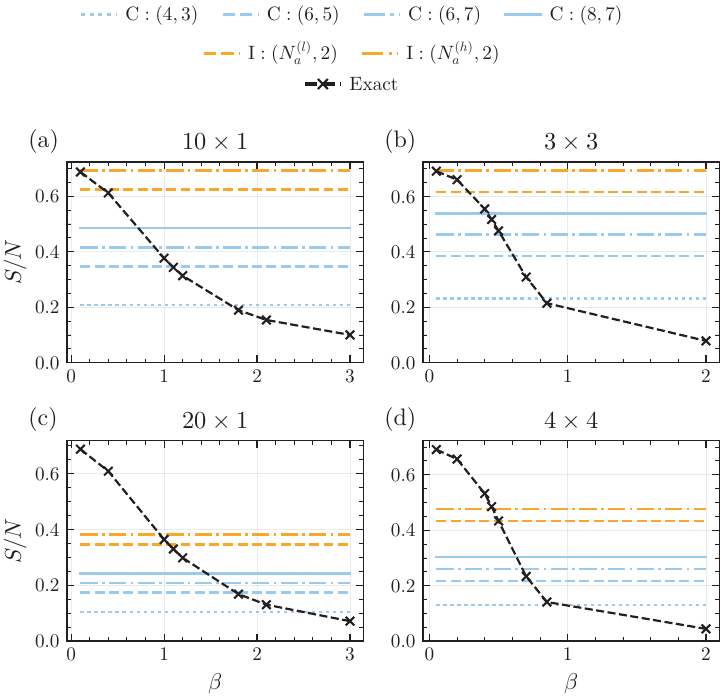}
  \caption{\revision{Entropy capacities of the contiguous and interleaved HEAs compared with the exact thermal entropy of the open-boundary TFIM with $J=1$ and $h=0.5$.
      Panels (a)--(d) correspond to $10\times1$, $3\times3$, $20\times1$, and $4\times4$ systems, respectively.
      Black crosses show the exact Gibbs-state entropy per spin, $S/N$, at inverse temperature $\beta$.
      Blue horizontal lines give the contiguous (C) entropy bound $\min(N_a,L)\ln2/N$, while orange lines give the interleaved (I) bound $N_a\ln2/N$; $N_a$ and $L$ denote the numbers of ancillas and layers.
      The lower/higher interleaved ancilla budgets $(N_a^{(l)},N_a^{(h)})$ are $(9,10)$ for $10\times1$, $(8,9)$ for $3\times3$, and $(10,11)$ for both larger systems.
  Interleaved $L=2$ and $L=3$ circuits with the same $N_a$ share one upper bound and are not drawn separately.}}
  \label{fig:entropy-resources}
\end{figure}

{\revcolor
  At any finite temperature, the exact Gibbs state of a finite system is full rank, so exact purification requires at least $N$ ancilla qubits.
  Our goal with $N_a<N$ is instead an approximate purification.
  Its quality depends in part on how strongly the thermal weights are concentrated in a low-dimensional subspace, as well as on the circuit's expressivity and its optimization.
}
At lower temperatures, the Boltzmann weights are heavily concentrated in the low-lying eigenstates.
As temperature increases, thermal fluctuations populate a much broader range of energy levels, making the state highly mixed.
{\revcolor
  This broadening can be quantified by the entropy-based effective rank, $r_{\rm eff}(\rho)=\exp[S(\rho)]$.
  For a density matrix $\rho$ of rank $r$, the effective rank satisfies $1\le r_{\mathrm{eff}}(\rho)\le r$, which need not be an integer and depends on the eigenvalue weights, not just on how many are nonzero.
  It equals one for a pure state and reaches the upper bound $r$ when $\rho$ is an equal mixture of $r$ orthogonal states.
  For a fixed Hamiltonian, the effective rank of a Gibbs state is nondecreasing with temperature and approaches $2^N$ at infinite temperature.
}

{\revcolor
  For any purification, the rank of the reduced system state is bounded by the lesser of the system and ancilla Hilbert-space dimensions:
  \begin{equation}
    \rank(\rho_S)\le2^{\min(N,N_a)}.
    \label{eq:generic-rank}
  \end{equation}
  Since $N_a \leq N$ in our setup, the ancilla Hilbert-space dimension places an immediate bound on every purification:
  \begin{equation}
    \rank(\rho_S)\le2^{N_a}.
    \label{eq:ancilla-rank}
  \end{equation}
  The contiguous HEA has an additional restriction: only one CNOT crosses the $S|A$ partition in each layer.
  A CNOT connecting the two registers has two conditional branches: the target is unchanged when the control is $0$, and flipped when it is $1$.
  Its operator is therefore a sum of two tensor products whose factors are mutually orthogonal on each register:
  \begin{equation}
    \mathrm{CNOT}_{a\to s}
    =|0\rangle\langle 0|_a\otimes I_s
    +|1\rangle\langle 1|_a\otimes X_s,
    \label{eq:cnot-schmidt}
  \end{equation}
  where $a \in A$ and $s \in S$ are the system and ancilla qubits connected by the CNOT.
  The operator Schmidt rank of the CNOT is therefore two.
  Applying it to a state with $r$ Schmidt terms produces at most $2r$ product terms, so the Schmidt rank of the resulting state can at most double.
  Starting from a product state, $L$ layers can consequently generate Schmidt rank no larger than $2^L$.
  Combining this layer restriction with the ancilla bound in Eq.~\eqref{eq:ancilla-rank} gives
  \begin{equation}
    \rank(\rho_S)\le2^{\min(N_a,L)}.
    \label{eq:contiguous-rank}
  \end{equation}
  Its entropy ceiling is therefore
  \begin{equation}
    S(\rho_S)\le S_{\rm cap}^{\rm cont}=\min(N_a,L)\ln2.
    \label{eq:contiguous-capacity}
  \end{equation}

  The interleaved HEA, on the other hand, couples each ancilla to a distinct system qubit within every layer.
  Unlike the contiguous circuit, which has only one CNOT connecting the two registers per layer, it provides $N_a$ separate links across the $S|A$ partition.
  With suitable single-qubit rotations in $U_R^{(l)}$ [cf.~Eq.~\eqref{eq:interleaved-rotations}], these CNOTs can create $N_a$ Bell pairs in a single layer.
  In this case, the ancilla register is maximally mixed, $\rho_A=I_A/2^{N_a}$, and the subsequent system-only gates leave its reduced state unchanged.
  The entropy ceiling is therefore
  \begin{equation}
    S(\rho_S)\le S_{\rm cap}^{\rm int}=N_a\ln2
    \label{eq:generic-capacity}
  \end{equation}
  for any $L\ge1$, rather than being additionally restricted by the layer number as in the contiguous HEA.
  Additional layers can change which states the circuit represents, but cannot increase this ceiling.

  Here $S_{\rm cap}$ denotes a formal upper bound, not the achieved entropy by a given circuit or a guarantee that the circuit saturates it.
  Both entropy ceilings also bound the effective rank: $r_{\rm eff}(\rho_S)\le\exp(S_{\rm cap})$.
  At $\beta\rightarrow0$, the target is maximally mixed.
  The largest squared Uhlmann fidelity attainable by a rank-$r$ state is then $r/2^N$, so a rank much smaller than $2^N$ precludes high global fidelity.
  At finite temperature, $S(\rho_\beta)\le S_{\rm cap}$ is necessary for exact representability but not sufficient: entropy is only one property of the target spectrum, and the circuit must also reproduce its eigenvalue distribution and correlations.
  Conversely, when the target entropy exceeds the circuit's capacity, exact preparation is impossible, but selected local observables may still be estimated accurately.
  This motivates our benchmarks in Sec.~\ref{sec:shared-benchmarks}.
}

{\revcolor
  Fig.~\ref{fig:entropy-resources} compares these entropy ceilings with the exact thermal entropy for the systems studied in detail in Sec.~\ref{sec:shared-benchmarks}.
  We compare four contiguous HEA configurations specified by $(N_a, L)$, and two interleaved HEA configurations with different ancilla budgets $N_a^{(l)}$ and $N_a^{(h)}$, representing lower and higher ancilla counts, respectively.
  The values of the ancilla budgets for different systems are given in Table~\ref{tab:shared-configurations}.
  Whenever a ceiling lies below the thermal curve, it means that no optimization within that circuit can reproduce the target entropy.
  Interleaving removes the additional layer-dependent restriction on entropy capacity in contiguous layouts, but the additional capacity gained by extending the ancilla register is useful only if the classical computation does not become prohibitively expensive.
  The MPS ordering is important here: a single bond separates the system and ancilla registers in the contiguous layout, whereas the interleaved partition crosses multiple bonds.
  We next examine how this difference affects the required bond dimensions and entropy-evaluation cost, before comparing the resulting thermal states in Sec.~\ref{sec:shared-benchmarks}.
}

\subsection{\revheading{Classical tensor-network cost}}
\label{sec:mps-cost}

{\revcolor
  An MPS bond separates the ordered qubits into a left and a right block.
  The bond dimension $D_b$ bounds the Schmidt rank across that cut.
  For contiguous registers, the $S|A$ partition coincides with a single such cut, so $\rank(\rho_S)\le D_{A|S}$ and $S(\rho_S)\le\ln D_{A|S}$.
  For interleaved registers, separating the system from the ancilla registers instead cuts several virtual bonds.
  Denoting this set of bonds by $\mathcal B_{S|A}$, we have
  \begin{equation}
    \rank(\rho_S)\le\min\left\{2^{N_a},\prod_{b\in\mathcal B_{S|A}}D_b\right\}.
    \label{eq:mps-register-rank}
  \end{equation}
  This product bound explains why the reduced-state rank can exceed every individual MPS bond dimension in the interleaved layout.
  For example, a product of adjacent ancilla--system Bell pairs, with any unpaired system qubits in $\ket{0}$, has maximum MPS bond dimension two, yet gives system entropy $N_a\ln2$.

  To propagate either circuit without truncation, the MPS must accommodate the Schmidt ranks generated across its individual cuts throughout the circuit.
  Counting the CNOTs that cross each cut gives a conservative upper bound on the required bond dimensions.
  Here we define the maximum bond dimension $D=\max_bD_b$ over all cuts $b$.
  Each CNOT crossing an MPS cut can at most double its Schmidt rank, by the operator decomposition in Eq.~\eqref{eq:cnot-schmidt}.
  Starting from a product state, the required bond dimensions therefore grow with the number of crossing gates.
  Every cut in the contiguous staircase is crossed by one CNOT per layer, giving the sufficient maximum bond dimension $D=2^L$.
  In the interleaved circuit, an individual MPS cut is crossed by at most one ancilla--system CNOT and one system--system CNOT per layer, giving the sufficient value $D=4^L$.
  Our calculations retain sufficient bond dimensions to propagate the circuits without truncation, so these numerical caps impose no additional restriction on the ansatz.
  Increasing them beyond the sufficient values would not enlarge the circuit's set of accessible states.
  On the other hand, a smaller cap would introduce additional numerical approximation errors.

  The classical cost is controlled by both the number of qubits and the bond dimensions needed during propagation.
  To compare the two layouts, we distinguish computation time from memory and account for three steps needed for MPS-assisted variational Gibbs-state preparation: propagating the purification circuit, computing its energy, and computing its entropy.
  The following estimates are sufficient time and storage bounds for standard tensor-network procedures at a fixed set of circuit parameters.
  Let $n=N+N_a$ be the number of qubits and $G$ the number of elementary gates.
  For each HEA considered here, $G = \mathcal O(nL)$, where $L$ is the number of layers.
  In each estimate, $D$ is the maximum bond dimension for the layout under consideration, which may differ between the two layouts.
  We denote computation time by $T$ and memory by $M$, with subscripts identifying the corresponding contributions.

  \emph{Circuit propagation.}
  Storing the purification as an MPS requires $M_{\mathrm{prop}}=\mathcal O(nD^2)$ memory~\cite{orusPracticalIntroductionTensor2014}.
  A standard two-site MPS gate update followed by matrix factorization takes $\mathcal O(D^3)$ time for fixed local Hilbert-space dimension~\cite{schollwockDensitymatrixRenormalizationGroup2011}.
  In both layouts, each gate acts within a fixed-size interval of the MPS ordering, giving total propagation time $T_{\mathrm{prop}}=\mathcal O(GD^3)$ without increasing the memory scaling.
  The MPS remains stored during the subsequent energy and entropy calculations; their memory bounds below count additional workspace.

  \emph{Energy computation.}
  The energy can be obtained by contracting the purification with an MPO representation of $H\otimes I_A$, without constructing $\rho_S$ explicitly.
  Let $W$ denote the maximum bond dimension of this MPO, distinct from the state bond dimension $D$.
  The MPO is stored in factorized form as $n$ local tensors, each with two physical indices of dimension two and two virtual indices of dimension at most $W$.
  Storing these local tensors densely therefore requires $M_{\mathrm{MPO}}=\mathcal O(nW^2)$ memory.
  A standard contraction takes $T_E=\mathcal O\!\left(nWD^3+nW^2D^2\right)$ time and $M_E=\mathcal O(WD^2)$ additional workspace, beyond the MPS and MPO storage~\cite{schollwockDensitymatrixRenormalizationGroup2011}.
  For the 1D nearest-neighbor chain Hamiltonians considered, $W$ can be chosen independent of system size; for the square-lattice TFIM in row-major order, a construction with $W=\mathcal O(w)$ is available, where $w$ is the lattice width.
  Inserting identity operators at ancilla positions does not require increasing $W$.

  \emph{Entropy computation.}
  In the contiguous layout, bringing the MPS into Schmidt form at the system--ancilla boundary straightforwardly yields the weights in Eq.~\eqref{eq:schmidt-entropy}.
  This gives the time bound $T_{S,\mathrm C}=\mathcal O(nD^3)$, with additional workspace bounded by $M_{S,\mathrm C}=\mathcal O(nD^2)$.
  In the interleaved layout, no single bond separates the registers.
  Direct entropy evaluation can instead proceed in three steps.
  First, construct $\rho_A=\Tr_S\ket{\psi}\!\bra{\psi}$ by contracting the system indices while leaving the ancilla indices open, using the standard tensor-network representation of a partial trace~\cite{schollwockDensitymatrixRenormalizationGroup2011}.
  Second, diagonalize this $2^{N_a}\times2^{N_a}$ matrix to obtain its eigenvalues $p_j$.
  Finally, evaluate $-\sum_jp_j\ln p_j$, which equals $S(\rho_S)$ due to the structural symmetry of the purification.

  The cost of constructing the reduced matrix depends on the contraction order~\cite{orusPracticalIntroductionTensor2014}.
  For the following bound, we split the MPS into left and right parts containing nearly equal numbers of ancillas, contract each part separately, and then join them across their connecting virtual indices.
  Unlike a factorized MPO, each contracted half is stored as a single dense tensor.
  A half containing $m$ ancillas retains $2^m$ ket configurations, $2^m$ bra configurations, and two virtual indices of dimension at most $D$, requiring at most $4^mD^2$ entries.
  Since $m\le\lceil N_a/2\rceil$, the partial-tensor memory is bounded by $M_{\mathrm{half}}=\mathcal O(2^{N_a}D^2)$.
  Building both partial tensors takes $T_{\mathrm{half}}=\mathcal O(n2^{N_a}D^3)$ time, and joining them into $\rho_A$ takes $T_{\mathrm{join}}=\mathcal O(4^{N_a}D^2)$ time.
  Dense diagonalization then takes $T_{\mathrm{diag}}=\mathcal O(8^{N_a})$ time and $M_{\mathrm{diag}}=\mathcal O(4^{N_a})$ memory, including the reduced matrix.
  The final entropy sum takes only $T_{\mathrm{sum}}=\mathcal O(2^{N_a})$ time and is subleading.
  Collecting these steps gives $T_{S,\mathrm I}=\mathcal O\!\left(n2^{N_a}D^3+4^{N_a}D^2+8^{N_a}\right)$.
  The additional workspace is bounded by $M_{S,\mathrm I}=\mathcal O(2^{N_a}D^2+4^{N_a})$, including the partial tensors and the dense matrix that coexist during joining.

  For either layout $X\in\{\mathrm C,\mathrm I\}$, adding the three time contributions gives the overall objective-evaluation time $T_X=T_{\mathrm{prop}}+T_E+T_{S,X}$.
  To bound peak memory, we count the persistent MPS and Hamiltonian storage, $M_{\mathrm{prop}}+M_{\mathrm{MPO}}$, together with the larger additional workspace, $\max\{M_E,M_{S,X}\}$, rather than summing the peaks of successive stages.
  For the contiguous HEA, the resulting bounds are
  \begin{subequations}
    \label{eq:contiguous-mps-cost}
    \begin{align}
      T_{\mathrm C}&=\mathcal O\!\left((G+nW)D^3+nW^2D^2\right),
      \label{eq:contiguous-mps-time}\\
      M_{\mathrm C}&=\mathcal O\!\left((n+W)D^2+nW^2\right).
      \label{eq:contiguous-mps-memory}
    \end{align}
  \end{subequations}
  For the interleaved HEA, they become
  \begin{subequations}
    \label{eq:interleaved-mps-cost}
    \begin{equation}
      \begin{split}
        T_{\mathrm I}=\mathcal O\!\bigl[(G+nW+n2^{N_a})D^3\\
        {}+(nW^2+4^{N_a})D^2+8^{N_a}\bigr],
      \end{split}
      \label{eq:interleaved-mps-time}
    \end{equation}
    \begin{equation}
      \begin{split}
        M_{\mathrm I}=\mathcal O\!\bigl[(n+W+2^{N_a})D^2 +nW^2+4^{N_a}\bigr].
      \end{split}
      \label{eq:interleaved-mps-memory}
    \end{equation}
  \end{subequations}

  In summary, compared to the contiguous HEA, interleaving changes both the sufficient bond dimensions and the entropy calculation.
  Even when its MPS bond dimensions remain small, the dense-matrix contributions $M_{\mathrm{diag}}$ and $T_{\mathrm{diag}}$ grow exponentially with $N_a$.
  There is therefore a trade-off between thermal-state expressivity and classical cost.
  Interleaving can provide much more entropy capacity at shallow depth, but each added ancilla exponentially increases the classical cost of evaluating the entropy.
  In Appendix~\ref{app:mps-resources} we report the chosen bond caps, achieved bond dimensions, as well as optimization and reconstruction times.
}

\section{Thermal-state numerical benchmarks}
\label{sec:shared-benchmarks}

{\revcolor
  For the numerical benchmarks, we minimize the dimensionless free energy per spin,
  \begin{equation}
    C_\beta(\rho_S)=\frac{\beta E(\rho_S)-S(\rho_S)}{N}.
    \label{eq:Cbeta}
  \end{equation}
  At fixed $N$ and positive $\beta$, this is a positive rescaling of $F_\beta$ and has the same minimizers.
  Dividing by $N$ also makes the objective convenient for comparing different system sizes.

  \begin{table}[t]
    \centering
    \revcolor
    \caption{\revision{Interleaved HEA configurations used in the numerical benchmarks.
        Three configurations are tested on each system, with two different ancilla budgets at $L=2$ and one $L=3$ circuit.
        $l/h$ denotes lower/higher ancilla budget within that system.
    Every system also has the four contiguous configurations $(4,3),(6,5),(6,7),(8,7)$.}}
    \label{tab:shared-configurations}
    \vspace{0.5em}
    \begin{tabular}{@{}l@{\quad}c@{\quad}c@{\quad}c@{}}
      \toprule
      System & $(N_a^{(l)},2)$ & $(N_a^{(h)},2)$ & $(N_a,3)$\\
      \midrule
      $10\times1$ & $(9,2)$ & $(10,2)$ & $(10,3)$\\
      $20\times1$ & $(10,2)$ & $(11,2)$ & $(10,3)$\\
      $3\times3$ & $(8,2)$ & $(9,2)$ & $(9,3)$\\
      $4\times4$ & $(10,2)$ & $(11,2)$ & $(10,3)$\\
      \bottomrule
    \end{tabular}
  \end{table}
}

We perform noiseless simulations of the MPS-assisted variational Gibbs state preparation algorithm for the TFIM on 1D chains with \revision{10 and 20} sites, and on 2D square lattices with \revision{$3\times3$ and $4\times4$} sites.
For square lattices, the Ising interaction in Eq.~\eqref{eq:tfim} is extended to all horizontal and vertical nearest-neighbor pairs, with open boundaries.
For all the simulations, we set the ferromagnetic coupling strength $J=1$ and the transverse field $h=0.5$.
{\revcolor
  We compare both HEA families with exact thermal references on these open systems.
  The one-dimensional inverse-temperature grid is
  $\beta_{\rm 1D}\in\{0.1,0.4,1.0,1.1,\allowbreak1.2,1.8,2.1,3.0\}$;
  the two-dimensional grid is
  $\beta_{\rm 2D}\in\{0.05,0.2,0.4,0.45,\allowbreak0.5,0.7,0.85,2.0\}$.
  These grids span high temperatures, where many energy levels are populated, through intermediate regimes to lower temperatures, where thermal excitations are increasingly suppressed.
  The physical meaning of the intermediate regime differs between one and two dimensions.

  The one-dimensional TFIM has no finite-temperature phase transition, but its bulk quasiparticle gap, $\Delta_{\rm 1D}=2|J-h|=1$, sets a characteristic thermal crossover scale~\cite{sachdevQuantumPhaseTransitions2011}.
  For the finite $10\times1$ and $20\times1$ chains, the corresponding excitation gaps are $\Delta_{\rm exc}\simeq1.11167$ and $1.02684$, respectively, where the excitation gap is the energy difference between the nearly generate ground-state doublet and the next excited state.
  The characteristic inverse-temperature scales are therefore $\beta_\Delta=1/\Delta_{\rm exc}\simeq0.89955$ and $0.97386$, respectively.
  At high temperatures, $\beta^{-1} = k_BT\gg\Delta_{\rm 1D}$, the system is highly mixed, with many populated energy levels contributing to the thermal state.
  When the thermal energy $\beta^{-1}$ is comparable to this gap, excitation populations can change appreciably with small temperature variations.
  At lower temperatures, the domain-wall excitations become dilute and spatial correlations extend over longer distances.
  We therefore sample the intermediate regime more closely at $\beta_{\rm 1D}=1.0,1.1,1.2$, around $\beta\Delta_{\rm 1D}\sim1$.
  This relation therefore identifies a coarse crossover scale for the 1D chains, but different observables can change most rapidly at different temperatures.

  In two dimensions, the infinite square lattice instead has a genuine thermal ordering transition.
  At zero transverse field, its critical inverse temperature is $\beta_cJ=\tfrac12\ln(1+\sqrt2)\simeq0.44069$; the weak field $h/J=0.5$ slightly suppresses the ordering temperature, placing the transition near $\beta_cJ\sim0.45$ in the published phase diagram~\cite{hesselmannThermalIsingTransitions2016}.
  The denser points $\beta_{\rm 2D}=0.4,0.45,0.5$ sample this intermediate-temperature region.
  Unlike the gap-based scale used for the 1D chains, this collective ordering transition is not determined by the inverse of the lowest excitation energy.
  Instead, it is a many-body effect that emerges only in the thermodynamic limit.
  Our finite open square lattices exhibit rounded, observable-dependent crossovers rather than a sharp transition.

  Regarding the circuit resources, we benchmark four contiguous HEA configurations with $(N_a,L)\in\{(4,3),(6,5),(6,7),(8,7)\}$.
  Three interleaved HEA configurations are also tested on each system, with two different ancilla budgets at $L=2$ and one $L=3$ circuit (Table~\ref{tab:shared-configurations}).
  All seven configurations per system are shown in the analyses below, giving 28 system--circuit combinations and 224 temperature-dependent state estimates in total for each observable.
  In all the figure legends throughout this section, C and I denote the contiguous and interleaved architectures.
  The symbols $N_a^{(l)}$ and $N_a^{(h)}$ distinguish the lower and higher interleaved ancilla counts within each system, rather than fixed budgets common to all systems.

  For each fixed system and circuit shape, we generate 20 independent random parameter vectors.
  Each optimization begins from one of these cold starts, and the same vectors are reused across the eight $\beta$ values.
  Different circuit configurations use independently generated vectors.
  During optimization, we propagate the purification circuits directly as MPSs using a custom JAX implementation~\cite{bradburyJAXComposableTransformations2018} and obtain objective gradients by automatic differentiation.
  We first use Adam to explore the objective landscape and then L-BFGS-B to refine the parameters.
  To check the optimized results independently, we construct the same circuits in Qiskit~\cite{javadiabhariQuantumComputingQiskit2024}, bind the saved parameter values, and simulate them as MPSs using Quimb~\cite{grayQuimbPythonPackage2018}.
  We recompute the energy, entropy, and objective through this independent implementation (Appendix~\ref{app:timing}) and select the reported state according to the lowest recomputed $C_\beta$.
  Only the selected state with the lowest $C_\beta$ is used for the observable comparisons below.
  Appendix~\ref{app:optimization-diagnostics} examines the variation across all 20 starts.
  In Appendix~\ref{app:gradients}, we also examine the scaling of objective gradients with system size, which is relevant to the trainability of the variational ansatz and serves as a finite-size check for barren plateaus~\cite{mccleanBarrenPlateausQuantum2018}.
  Finally, exact references come from a free-fermion solution for the open chains and complete diagonalization for the square lattices.
  Appendix~\ref{app:reference-construction} describes the exact thermal and ground-state references in more detail.

  We assess the results at three levels.
  The objective error, $\Delta C_\beta=C_\beta-C_\beta^{\rm ref}$, measures how well the variational state approximates the target thermal state from a global perspective.
  A nonzero value can arise from limited circuit expressivity, including insufficient entropy capacity discussed in Sec.~\ref{sec:rank-entropy-capacity}, or incomplete optimization, or both.
  Energy and entropy densities, fluctuation estimators, and spatial correlations reveal which physical properties are nevertheless reproduced.
  Finally, the variation across initializations shows how consistently optimization finds a good state.
  Keeping these levels distinct helps explain why an accurate observable need not imply a globally accurate thermal state, or a particularly good run does not imply reliable optimization.
}

\subsection{\revheading{Global objective, energy, and entropy}}

\begin{figure*}[t]
  \centering
  \figurefile[\textwidth]{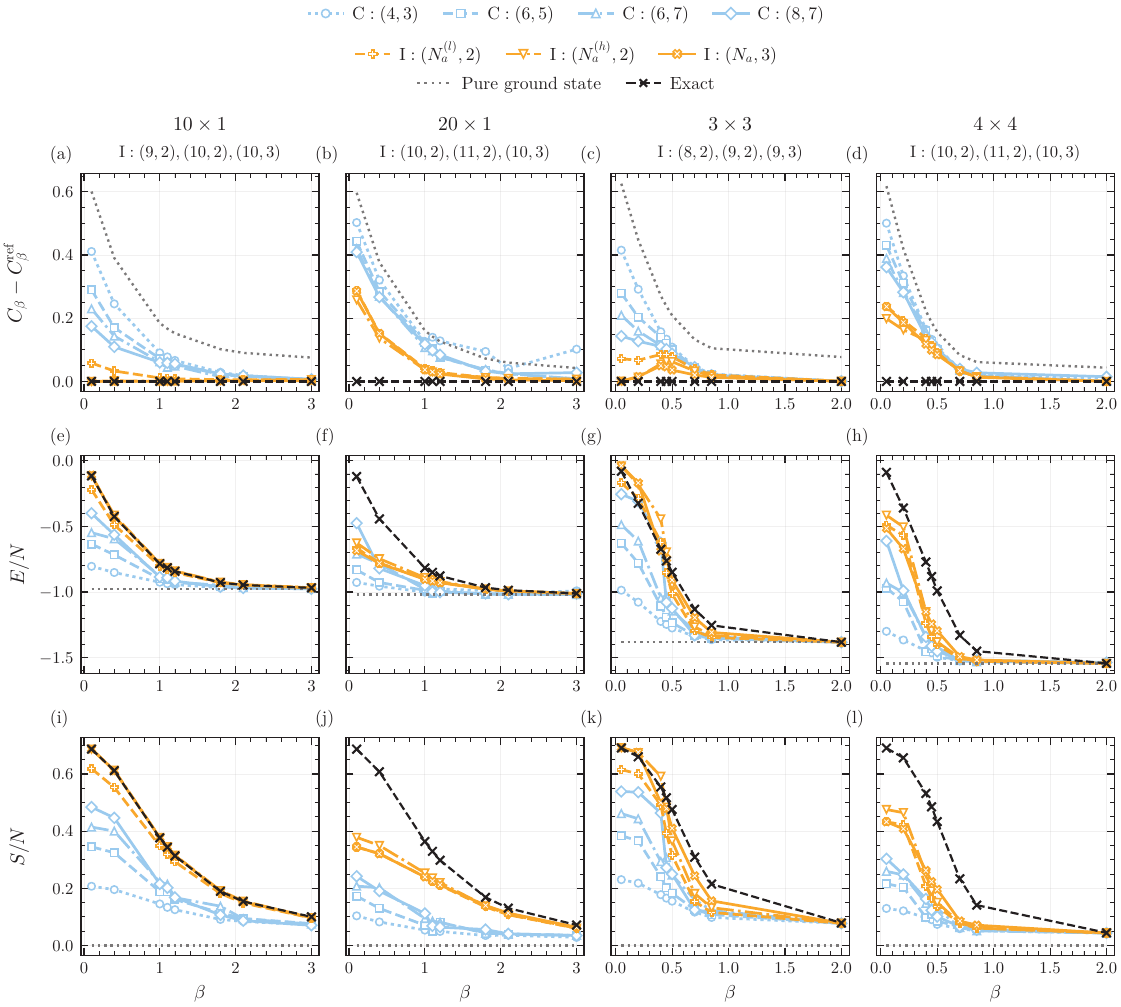}
  \caption{\normalcolor\revision{Thermal-state estimates from the contiguous and interleaved HEAs for the open-boundary TFIM with $J=1$ and $h=0.5$, at various inverse temperatures $\beta$.
      Columns correspond to $10\times1$, $20\times1$, $3\times3$, and $4\times4$ systems, respectively.
      From top to bottom, the rows show the objective error $C_\beta-C_\beta^{\rm ref}$ in panels (a)--(d), energy per spin $E/N$ in (e)--(h), and entropy per spin $S/N$ in (i)--(l).
      Blue and orange curves denote contiguous (C) and interleaved (I) circuits; legend pairs give the numbers of ancillas and layers, $(N_a,L)$.
      Superscripts $l/h$ distinguish lower/higher interleaved ancilla budgets, with numerical configurations given above each column.
    Black dashed curves with crosses show the exact thermal references, while gray dotted curves show the predictions based on the pure ground state.}
    Each variational data point is the best result from the 20 optimization runs with different initial parameters.
  \revision{The same state, selected by the lowest independently recomputed $C_\beta$, supplies all three quantities.}}
  \label{fig:shared-global}
\end{figure*}

{\revcolor
  Fig.~\ref{fig:shared-global} shows how the energy--entropy balance affects the quality of the prepared state.
  The free-energy objective depends on both the energy and entropy of the trial state: at high temperature, the target entropy approaches $N\ln2$, and failing to reproduce it incurs a large objective penalty even when the energy is reasonable.
  Because entropy enters $F_\beta=E-S/\beta$ with a minus sign, lowering the energy below the exact value can partly compensate for missing entropy.
  This compensation is limited, however, because energy and entropy must come from the same physical state and cannot be adjusted independently.
  At finite positive $\beta$, only the Gibbs state attains the minimum free energy in Eq.~\eqref{eq:gibbs-variational}.
  An entropy-deficient state therefore cannot reproduce the exact free energy by lowering its energy enough to cancel the entropy deficit completely.
  The optimizer may nevertheless favor a state with lower energy and entropy than the target, without obtaining the correct Gibbs state.
  This is most apparent for the $20\times 1$ and $4\times 4$ systems at high and intermediate temperatures, where the energy and entropy estimates of all circuit configurations fall below the exact values, and the resulting objective errors are positive, as seen in the second and fourth columns of Fig.~\ref{fig:shared-global}.

  Between the two HEA families, the rank-limited contiguous circuits have their largest objective gaps at high temperatures, even on the smaller $10\times1$ and $3\times3$ systems, where the target entropy exceeds the circuits' ceilings [cf.~Figs.~\ref{fig:entropy-resources}(a)--(b)].
  The tested interleaved candidates generally reduce these gaps, consistent with their greater entropy capacity.
  On the smaller systems, the interleaved circuits tested here can reasonably reproduce the target energy and entropy across the entire temperature range, with small objective errors.
  However, even the interleaved circuit with the largest ancilla budget ($N_a = 11$) exhibits a large objective gap at high temperature for the larger $20\times1$ and $4\times4$ systems.
  Focusing on the entropy estimates in Figs.~\ref{fig:shared-global}(j) and (l), we note that the interleaved predictions can still be below the exact values at intermediate temperatures where the target entropy falls below the circuits' ceilings [cf.~Figs.~\ref{fig:entropy-resources}(c)--(d)].
  This indicates that entropy capacity is merely one factor in determining the quality of the variational state, and other factors, such as the circuit expressivity and the optimization landscape, also play important roles.

  Adding one ancilla to the interleaved circuit at fixed $L=2$ lowers the best $C_\beta$ at all 32 sampled system--temperature combinations, with the largest improvements seen at high temperature.
  The selected states have both greater entropy and higher (less negative) energy.
  The entropy gain more than compensates for the energy increase in the objective, so the overall $C_\beta$ is lower.
  However, it is worth noting that besides raising the entropy ceiling, the increasing number of ancillas also adds variational parameters and redistributes the system--ancilla couplings.
  The observed improvement therefore reflects these changes together, rather than an isolated increase in entropy capacity.
  On the other hand, increasing the number of layers from two to three at fixed $N_a$ in the interleaved circuit does not lead to a significant difference in the selected objective in most cases.

  Finally, the exact pure-ground-state curves, shown as the gray dotted lines in Fig.~\ref{fig:shared-global}, provide a useful second comparison.
  With zero entropy and the lowest possible energy $E_0$, the ground state gives the best pure-state objective, $C_\beta^{\rm GS}=\beta E_0/N$.
  In the ferromagnetic regime studied here, the interaction favors two opposite longitudinal orientations, all spins up or all spins down, which are related by global spin inversion.
  On a finite lattice, the transverse field dresses these orientations with quantum fluctuations and allows weak tunneling between them, giving an even-parity ground state and a nearby odd-parity partner, which are separated by a small energy splitting.
  Here parity is the eigenvalue of the spin-inversion operator $\prod_i X_i$.
  At temperatures low enough to suppress higher excitations but still large compared with the small splitting within this doublet, the Gibbs state is a nearly equal incoherent mixture of these two states.
  Its entropy is then approximately $\ln2$, which the pure-ground-state reference misses despite having a very similar energy.
  As $\beta\rightarrow\infty$ at fixed finite size, all excited-state populations get suppressed, and the Gibbs state approaches the unique ground state with vanishing entropy.
  The distinction between the global states does not, by itself, establish an advantage for thermal-state preparation at low temperature.
  To assess its practical benefit, we further compare the observable errors of the prepared thermal states with those of the pure-ground-state reference in the following subsections.
}

\subsection{\revheading{Thermal fluctuations}}

\begin{figure*}[t]
  \centering
  \figurefile[\textwidth]{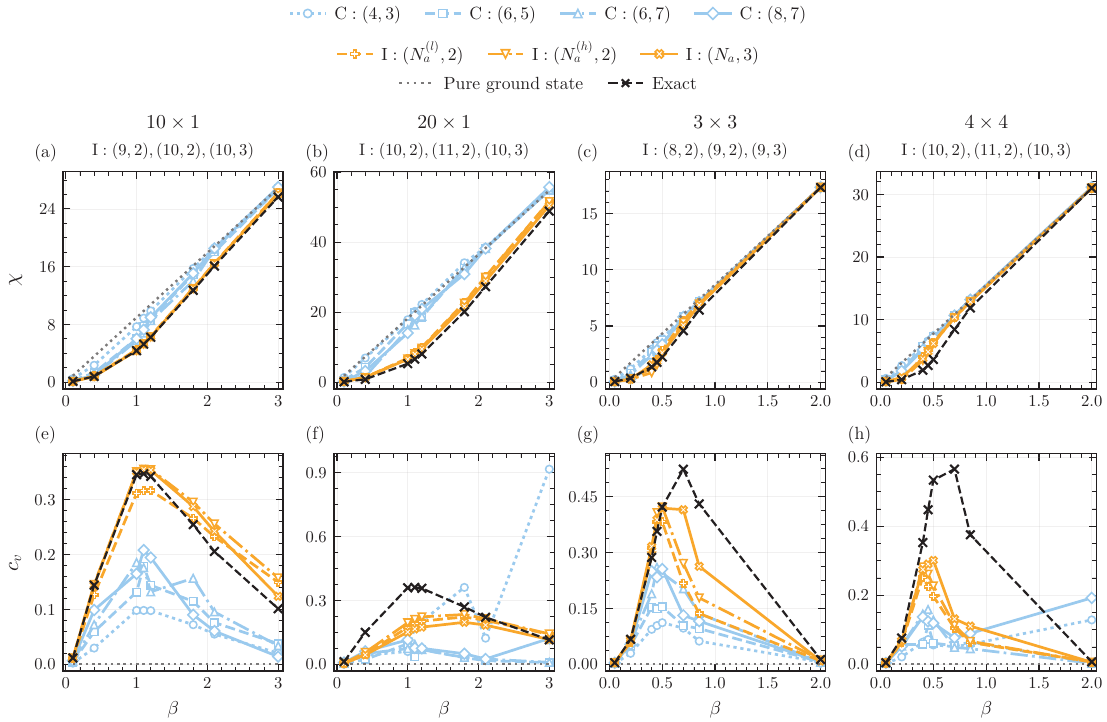}
  \caption{\normalcolor\revision{Thermal estimates of magnetization and energy fluctuations for the open-boundary TFIM with $J=1$ and $h=0.5$.
      Panels (a)--(d) show the magnetization-fluctuation estimator $\chi$, and panels (e)--(h) show the energy-fluctuation estimator $c_v$.
      Columns correspond to $10\times1$, $20\times1$, $3\times3$, and $4\times4$ systems, respectively.
      Blue and orange curves represent contiguous (C) and interleaved (I) HEAs with ancilla and layer counts $(N_a,L)$; $l/h$ label the lower/higher interleaved ancilla budgets, and column headings give the explicit interleaved configurations.
    Black dashed curves with crosses show exact thermal results, and gray dotted curves show the same estimators evaluated in the exact pure ground state.}
    Each variational data point is the best result from the 20 optimization runs with different initial parameters.
  }
  \label{fig:shared-observables}
\end{figure*}

{\revcolor
  Energy is the first moment of the state's energy distribution, but thermal fluctuations depend on second moments as well.
  We characterize thermal fluctuations using the magnetization-fluctuation estimator $\chi$ and the energy-fluctuation estimator $c_v$, defined as
  \begin{align}
    \chi&=\frac{\beta}{N}
    \left(\langle M^2\rangle-\langle M\rangle^2\right),\label{eq:chi}\\
    c_v&=\frac{\beta^2}{N}
    \left(\langle H^2\rangle-\langle H\rangle^2\right).
    \label{eq:cv}
  \end{align}
  Here $\langle O\rangle=\Tr(\rho_S O)$ is the expectation in the state being evaluated: $\rho_\beta$ for the thermal reference and the selected $\rho_S(\boldsymbol\theta)$ for a variational candidate.
  $M=\sum_{i=1}^{N}Z_i$ is the total longitudinal magnetization operator.
  The familiar fluctuation notation requires some care in a quantum system.
  Since $[M,H]\ne0$ due to the presence of the transverse field, $\chi$ is not exactly the static magnetic susceptibility.
  The identification holds only in the classical limit, where the Hamiltonian is diagonal in the $Z$ basis.
  Moreover, for an exact Gibbs state, $c_v$ corresponds to the specific heat per site.
  For a variational state, it instead measures energy fluctuations and does not necessarily equal $\partial(E_{\rm var}/N)/\partial T$, where $E_{\rm var} = \Tr(\rho_S(\boldsymbol\theta) H)$ is the energy along the optimized variational family.
}

{\revcolor
  Fig.~\ref{fig:shared-observables} illustrates the different information contained in these estimators.
  The upper row compares $\chi$ across the four systems, while the lower row shows the corresponding $c_v$ estimates.
  Starting with $\chi$, the interleaved HEA substantially improves the 1D results: all three interleaved candidates are more accurate than the four contiguous candidates at every sampled temperature for $10\times1$, and at all but the smallest $\beta$ for $20\times1$.
  At that highest-temperature point for $20\times1$, the best contiguous estimate is comparable to the interleaved estimates.
  The improvements for the square lattices are slightly less uniform.
  For example, the contiguous $(8,7)$ circuit gives a more accurate $\chi$ than any of the interleaved candidates for $3\times3$ at $\beta=0.2,0.4$, despite its larger objective error; this small difference is resolved in the absolute-error comparison in Fig.~\ref{fig:ground-state-errors}(c) (Appendix~\ref{app:ground-state-observables}).
  In the larger $4\times4$ system, increasing the interleaved ancilla budget from 10 to 11 at $L=2$ generally improves the $\chi$ estimates [cf.~Fig.~\ref{fig:ground-state-errors}(d)], but appreciable deviations remain at intermediate temperatures.

  Comparing Figs.~\ref{fig:shared-global} and ~\ref{fig:shared-observables} shows that substantial improvements in magnetization fluctuations can be present despite persistent errors in the energy.
  For the $20\times1$ chain at high and intermediate temperatures, the interleaved circuits improve $\chi$ considerably over the contiguous circuits, although their energies remain below the exact thermal values.
  These observables depend on different combinations of the state's correlations: the energy involves nearest-neighbor longitudinal correlations and single-site transverse magnetizations, whereas $\chi$ sums connected longitudinal correlations over all spin pairs.
  Their errors therefore need not track one another.
  Moreover, agreement in this sum does not guarantee an accurate spatial correlation profile, which we examine in Sec.~\ref{sec:correlations}.

  Turning to $c_v$, the exact references have their largest values on the sampled grids at $\beta=1.1$ for both chains and at $\beta=0.7$ for both square lattices.
  The interleaved circuits reproduce the peak well for $10\times1$, but the larger chain and square-lattice results show more pronounced distortions of the energy-fluctuation profile.
  Increasing the interleaved ancilla budget from 10 to 11 improves $c_v$ at seven of the eight temperatures for each larger system, with exceptions at $\beta=3$ for $20\times1$ and $\beta=0.2$ for $4\times4$.
  Thus the greater entropy capacity and lower selected objective do not guarantee improvement in every observable.

  The absolute and relative errors in Figs.~\ref{fig:ground-state-errors} and~\ref{fig:relative-observable-errors} (Appendix~\ref{app:ground-state-observables}) help interpret these temperature-dependent deviations.
  The explicit factors of $\beta$ and $\beta^2$ in Eqs.~\eqref{eq:chi} and~\eqref{eq:cv} suppress absolute errors in $\chi$ and $c_v$ at high temperature, even when the underlying variances are inaccurate.
  Dividing by the corresponding exact thermal values removes these factors and measures the fractional errors in the variances themselves.
  For the interleaved $20\times1$ circuits, the relative error in $\chi$ decreases on cooling, although its absolute error initially grows.
  By contrast, both observables retain substantial relative errors at intermediate temperatures for the interleaved $4\times4$ circuits, so those discrepancies cannot be explained solely by the temperature prefactors.
  The relative accuracy of the two estimators also changes with temperature.
  For the interleaved $4\times4$ circuit with $(N_a,L)=(11,2)$, the relative error in $c_v$ is smaller than that in $\chi$ at $\beta=0.4$, whereas the opposite holds at $\beta=0.85$ [cf.~Fig.~\ref{fig:relative-observable-errors}(d) and (h)].
  Thus the more conspicuous distortions of the $c_v$ profile in Fig.~\ref{fig:shared-observables} do not imply that its fractional error is consistently larger than that of $\chi$.

  Physically, $c_v$ probes the width of the energy distribution, not just its mean.
  Measured relative to the ground-state energy, an excitation's contribution is weighted by its energy in the first moment but by its squared energy in the second moment.
  Incorrect excitation populations can therefore distort the variance even when the mean energy is reasonably accurate.
  When the target variance is small, errors in the approximate state's two energy moments can produce a large fractional error in their difference.
  By contrast, global spin-flip symmetry gives $\langle M\rangle_{\rho_\beta}=0$ for the finite TFIM, while $\langle M^2\rangle_{\rho_\beta}$ can remain large when ferromagnetic correlations are strong.
  For variational states with a small mean magnetization, $\chi$ therefore avoids the subtraction of two large, nearly equal moments, although errors in the second moment can still limit its accuracy.
  These different sensitivities help explain why reproducing $\chi$ does not ensure an accurate $c_v$.
  Indeed, in the interleaved results on the square lattices, the largest absolute and relative errors in $c_v$ occur at or adjacent to the sampled maximum of the exact specific heat.
  This suggests that reproducing energy fluctuations is particularly challenging near that peak for these systems and circuit configurations.
  The correspondence is less consistent for the chains, where the error maxima can occur away from the $c_v$ peak and depend on whether absolute or relative errors are considered [cf.~Figs.~\ref{fig:ground-state-errors} and~\ref{fig:relative-observable-errors}].
  Magnetization-fluctuation errors also do not always peak in the same region.
  Thus, the specific-heat maximum identifies a challenging regime in some of our benchmarks, but not a universal point of greatest state-preparation difficulty.
}

{\revcolor
  Finally, the pure-ground-state reference in Fig.~\ref{fig:shared-observables} tests whether thermal preparation improves the observables over a zero-temperature approximation.
  It has zero energy variance and hence $c_{v,{\rm GS}}=0$, but a nonzero magnetization variance and a $\beta$-dependent $\chi_{\rm GS}$.
  Therefore, the ground state alone cannot reproduce finite thermal energy fluctuations, but it can give a good estimate of the magnetization fluctuations at low temperature.
  However, its small absolute $\chi$ error at the highest temperatures should not be mistaken for accurate thermal fluctuations: the relative-error comparison in Fig.~\ref{fig:relative-observable-errors}(a)--(d) shows large fractional discrepancies there.
  Every displayed interleaved candidate improves the $\chi$ estimates over this reference at all sampled temperatures for the 1D chains.
  The $(6, 7)$ and $(8, 7)$ contiguous circuits also improve $\chi$ over the ground state at intermediate temperatures, but the improvements are more modest than those of the interleaved circuits.
  For the square lattices, all three interleaved candidates also improve $c_v$ throughout and $\chi$ at every point except the coldest, $\beta=2$.
  At that point the pure ground state gives a comparable and even slightly more accurate $\chi$ than any of the displayed variational states.
  Thermal preparation therefore offers useful improvements over much of the sampled range, particularly at intermediate temperatures, but the ground state remains competitive for selected low-temperature observables.
}

\subsection{Spatial correlations}
\label{sec:correlations}

\begin{figure}[t]
  \centering
  \figurefile[\linewidth]{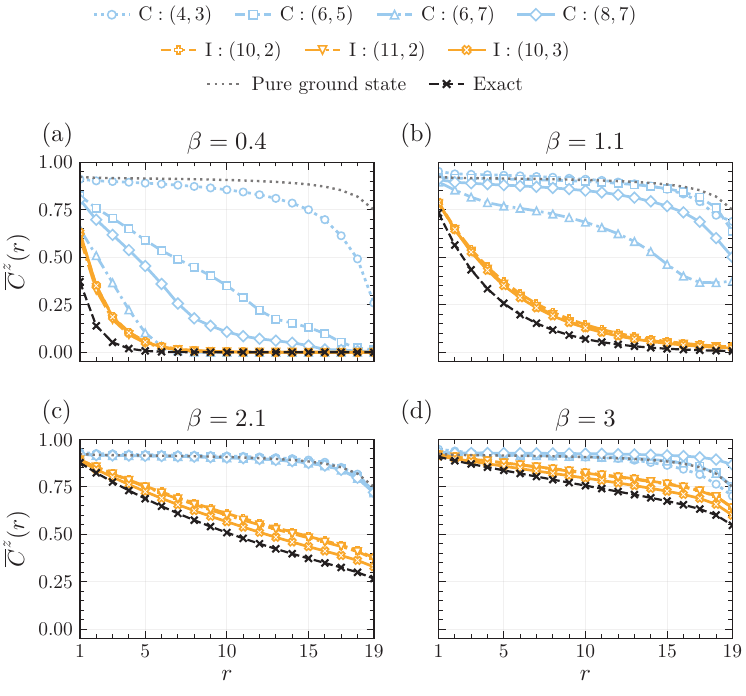}
  \caption{\revision{Thermal estimates of distance-averaged longitudinal correlations $\overline C^z(r)$ for the 20-spin open TFIM chain with $J=1$ and $h=0.5$.
      Panels (a)--(d) correspond to inverse temperatures $\beta=0.4,1.1,2.1,3.0$, respectively.
      Each point shows the mean connected correlation $\langle Z_iZ_j\rangle-\langle Z_i\rangle\langle Z_j\rangle$ over the $20-r$ pairs separated by $r=|i-j|$, while on-site terms are omitted.
      Blue and orange curves represent contiguous (C) and interleaved (I) HEAs, with the ancilla and layer counts $(N_a,L)$ specified in the legend.
      Black dashed curves with crosses and gray dotted curves show the exact thermal and pure-ground-state correlations, respectively.
  Each variational data point is the best result from 20 independent optimization runs with different initial parameters.}}
  \label{fig:shared-correlations}
\end{figure}

\begin{figure}[t]
  \centering
  \figurefile[\linewidth]{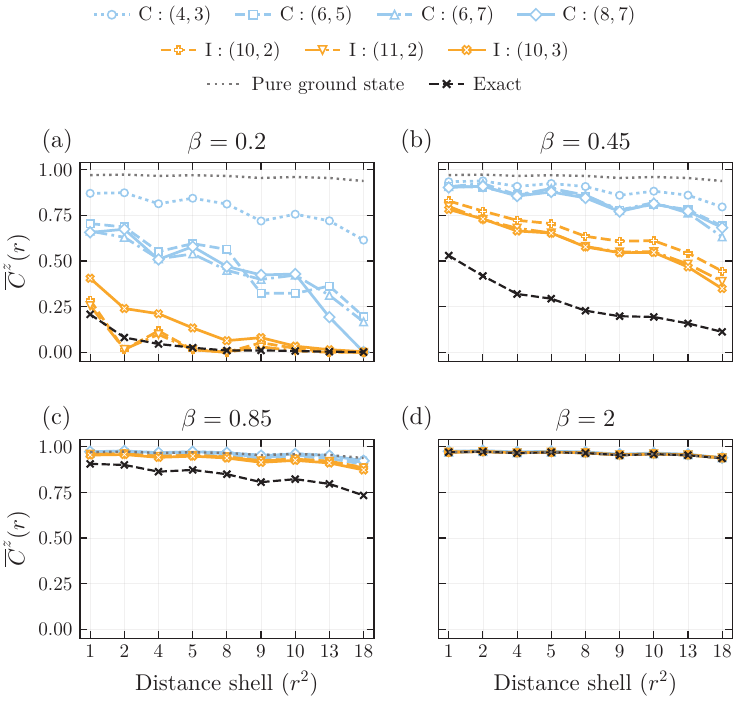}
  \caption{\revision{Thermal estimates of radially averaged longitudinal correlations $\overline C^z(r)$ for the open $4\times4$ TFIM with $J=1$ and $h=0.5$.
      Panels (a)--(d) correspond to inverse temperatures $\beta=0.2,0.45,0.85,2.0$, respectively.
      Each point averages the connected correlation $\langle Z_iZ_j\rangle-\langle Z_i\rangle\langle Z_j\rangle$ over all distinct unordered site pairs at Euclidean distance $r$.
      The nine distance shells are equally spaced for presentation and labeled by $r^2=1,2,4,5,8,9,10,13,18$ on the horizontal axis.
      Blue and orange curves represent contiguous (C) and interleaved (I) HEAs, with the ancilla and layer counts $(N_a,L)$ specified in the legend.
      Black dashed curves with crosses and gray dotted curves show the exact thermal and pure-ground-state references, respectively.
  Each variational data point is the best result from 20 independent optimization runs with different initial parameters.}}
  \label{fig:shared-correlations-4x4}
\end{figure}

To examine the spatial information behind the fluctuation estimates, we consider connected longitudinal correlations,
\begin{equation}
  C^z_{ij}=\langle Z_iZ_j\rangle-\langle Z_i\rangle\langle Z_j\rangle.
  \label{eq:connected-correlation}
\end{equation}
Two-point correlations such as $C^z_{ij}$ play a foundational role in many-body physics, as they capture how fluctuations at different sites are correlated and reflect the spatial structure of a thermal state.

{\revcolor

  Expanding the magnetization variance in Eq.~\eqref{eq:chi} gives the relation between $\chi$ and the connected correlations:
  \begin{equation}
    \chi=\frac{\beta}{N}\sum_{i,j}C^z_{ij}.
    \label{eq:chi-correlations}
  \end{equation}
  Unlike the energy of the local Hamiltonian, this sum includes correlations at every separation.
  Errors in the long-distance profile can therefore accumulate in $\chi$ even when nearest-neighbor terms are accurate.
  Conversely, errors of opposite sign at different locations can partially cancel in the sum.
  To examine the spatial information that is lost in the integrated quantity $\chi$, we average the connected correlations over all physical site pairs at a given separation.
  For an open chain, we define
  \begin{equation}
    \overline{C}^{z}(r)=\frac{1}{N-r}\sum_{i=1}^{N-r}C^z_{i,i+r},
    \label{eq:chain-averaged-correlation}
  \end{equation}
  for separation $r=1,\ldots,N-1$ in lattice-spacing units.
  This average includes all $N-r$ pairs at that separation, including pairs involving boundary sites, without assuming translation invariance.
  The profile therefore provides an summary of the spatial correlations, but can hide errors at individual sites, and fewer pairs contribute at larger separations.

  In the 1D ferromagnetic regime studied here, cooling makes thermally excited domain walls---boundaries between regions of opposite longitudinal magnetization---increasingly dilute.
  Spins farther apart are then less likely to be separated by such a boundary and can remain correlated over longer distances.
  The infinite chain nevertheless has a finite correlation length at every nonzero temperature, which is the characteristic distance over which correlations decay~\cite{sachdevQuantumPhaseTransitions2011}.
  On a finite chain, this length can become comparable to the system size, so correlations remain appreciable even between distant spins, despite the absence of a finite-temperature phase transition.
  Open boundaries can also modify this profile, as spins near the edges have fewer neighbors and are less likely to be separated by a domain wall.
  At fixed finite size, the Gibbs state approaches the unique ground state only as $\beta\to\infty$.
  Therefore, appreciable thermal changes in the spatial correlations can remain at the finite temperatures considered here.

  Fig.~\ref{fig:shared-correlations} illustrates this crossover for the $20\times1$ chain.
  The exact thermal correlations decay rapidly with separation in the warmest panel ($\beta = 0.4$), whereas cooling produces increasingly strong correlations even between widely separated spins.
  The contiguous circuits generally overestimate both the magnitude and spatial range of these correlations, with several profiles remaining much closer to the strongly correlated ground-state curve than to the thermal reference.
  The interleaved circuits reproduce the temperature-dependent decay substantially better, although their correlations also tend to lie above the exact values.
  In particular, their profiles retain the gradual thermal decay in the cooler panels, where the contiguous curves are comparatively flat through much of the chain.
  Thus the improved magnetization-fluctuation estimates discussed in the previous subsection are indeed accompanied by a more faithful spatial profile, rather than agreement only in the sum.

  Overall, differences among the interleaved curves are smaller than the separation between the two families.
  Increasing the $L=2$ ancilla budget gives a modest improvement, while the deeper $(10,3)$ circuit is visibly closer to the thermal reference at longer distances in the cooler panels.
  Additional layers can therefore improve spatial correlations without raising the entropy ceiling, which implies that the entropy ceiling alone does not determine observable accuracy.
  The pure-ground-state profile remains more strongly correlated than the thermal reference throughout the plotted range, especially at high and intermediate temperatures.
  Its contrast with the interleaved results shows why thermal preparation remains useful at the cooler temperatures, even when both the target and ground state have strong short-distance correlations.

  On the square lattice, we instead use radial averaging.
  That is, we average the correlations over all physical site pairs at the same Euclidean distance:
  \begin{equation}
    \overline{C}^{z}(r)=\frac{1}{n(r)}
    \sum_{(i,j)\in\mathcal P(r)}C^z_{ij}.
    \label{eq:square-averaged-correlation}
  \end{equation}
  Here $\mathbf r_i$ is the physical lattice position of site $i$, $\mathcal P(r)$ is the set of unordered pairs $i<j$ with $|\mathbf r_j-\mathbf r_i|=r$, and $n(r)$ is the number of such pairs.
  This average includes diagonal separations as well as pairs along the horizontal and vertical axes.
  For readability, Fig.~\ref{fig:shared-correlations-4x4} places the nine shells of the open $4\times4$ square at equal horizontal intervals on the x-axis and labels them by their physical $r^2$ values.
  We note that such radial averaging can nonetheless conceal directional and site-dependent errors.

  Unlike the 1D case, the infinite square-lattice TFIM in this ferromagnetic regime supports a genuine finite-temperature phase transition~\cite{hesselmannThermalIsingTransitions2016}.
  At sufficiently low temperature, collective order survives at nonzero temperature in the thermodynamic limit.
  The small open square studied here instead exhibits a rounded crossover from weak, short-ranged correlations to strong correlations across the lattice as the temperature is lowered.
  Fig.~\ref{fig:shared-correlations-4x4} follows this evolution with increasing $\beta$.

  In the warmest square-lattice panel, the interleaved profiles are much closer to the weakly correlated thermal reference than either the contiguous profiles or the pure-ground-state curve.
  Some interleaved estimates lie on opposite sides of the reference at different distance shells, rather than uniformly overestimating the correlations as in the 1D case.
  On cooling, the difference between the two families becomes smaller as all profiles become strongly correlated.
  In the coldest panel, the variational, thermal, and ground-state curves nearly coincide, so these correlations no longer demonstrate a clear benefit over the ground-state approximation.
  Agreement in this observable does not imply that the thermal state has become pure: both members of the low-energy doublet discussed previously can remain populated while giving similar longitudinal correlations.

  At intermediate temperatures, however, appreciable discrepancies remain even for the interleaved candidates, particularly in the crossover region sampled at $\beta=0.45$.
  Compared with the more faithful 1D profiles in Fig.~\ref{fig:shared-correlations}, these results suggest that capturing the intermediate-temperature correlations on the square lattice is more demanding for the tested MPS-ordered circuits.
  A plausible explanation arises from the spatial structure of thermal fluctuations, rather than the entropy ceiling alone.
  Near the phase crossover, domains of different sizes and shapes contribute to correlations across both lattice directions.
  As a result, reproducing their relative statistical weights can be more demanding than producing weak correlations or strong collective alignment.
  Both HEAs nevertheless apply their system entangling gates along a one-dimensional sequence, rather than along all neighboring pairs of the square lattice.
  Interleaving raises the available entropy capacity, but adding ancillas does not change this 1D pattern of entangling gates.
  With few layers, the circuit may therefore reproduce strong collective alignment more readily than the thermal distribution of spatial fluctuations, consistent with the excessive correlations observed here.
  This is a possible restriction of the circuit family, rather than one of the MPS simulation, because all lattice interactions are included, and the circuits are propagated without truncation (see the bond caps and truncation checks in Appendix~\ref{app:bond-dimensions}).
  Therefore, this comparison identifies a practical accuracy limitation for the tested finite 2D systems.
  Ansatzes with an entangling structure more closely aligned with the lattice geometry may be needed to improve the intermediate-temperature correlations, which is a promising direction for future work.
  However, this would also increase the number of long-range gates in the circuit and therefore the classical simulation cost, which is a trade-off that needs to be considered in designing more effective ansatzes for higher-dimensional systems.
}

\section{Conclusion and outlook}
\label{sec:conclusion}

In this work, we performed a systematic study of variational Gibbs-state preparation that leverages MPS techniques to classically simulate and optimize the ansatz circuits.
\revision{MPS assistance offers a route beyond full-statevector simulation when both circuit propagation and entropy evaluation remain tractable.
Our benchmarks examine how the choice of ansatz affects these costs as well as the quality of the resulting thermal state.}
As an initial ansatz comparison, we investigated the performance of two distinct ansatz designs, the TFDA~\cite{wuVariationalThermalQuantum2019,zhuGenerationThermofieldDouble2020} and the HEA~\cite{wangVariationalQuantumGibbs2021}, on small 1D TFIM and XXZ systems with up to 6 spins using statevectors.
Within the tested circuits and temperature range, the contiguous HEA performs better at low temperatures, while the TFDA performs better at high temperatures.

\revision{The contiguous HEA's high-temperature limitation motivates the interleaved architecture, which removes its one-bit-per-layer entropy restriction while retaining an entropy capacity bound imposed by the number of ancillas.
  This gain in capacity changes the classical resource requirements.
  As system size grows, the contiguous HEA has to increase its ancilla qubit count and depth at the same time to keep up with the increasing target entropy, which can be costly for both classical simulation and quantum hardware.
  In the MPS simulation, the maximum bond dimension needed for exact propagation grows exponentially with the number of layers, becoming one of the main bottlenecks for the contiguous structure.
  The interleaved HEA, on the other hand, can increase its ancilla count at fixed depth to raise its entropy capacity.
  However, while contiguous registers allow entropy evaluation at a single MPS bond, the interleaved entropy calculation needs to construct and diagonalize a reduced matrix of dimension $2^{N_a}$.
The additional entropy capacity of interleaving at shallow depths therefore comes at a computational price: in the entropy-evaluation procedure used here, the classical costs grow exponentially with the ancilla number.}

Entropy capacity alone, however, does not determine the quality of the prepared Gibbs state.
In the larger MPS-based simulations, we estimated various thermal observables, including energy density, entropy density, magnetization and energy fluctuations, and two-point correlations.
\revision{On the tested $10\times1$, $20\times1$, $3\times3$, and $4\times4$ TFIM systems, the tested interleaved circuits generally reduce the selected objective errors and improve many observables.
  Adding one ancilla at fixed $L=2$ lowers the best objective at every sampled system--temperature point, without always improving every individual observable.
  The energy-fluctuation profiles and distance-resolved correlations reveal discrepancies that are not apparent from agreement in the energy or total magnetization fluctuations alone.
  For the larger systems, the interleaved circuits reproduce the temperature dependence of magnetization fluctuations more faithfully than the energy-fluctuation profile.
  The spatial correlations also remain imperfect, particularly near the square-lattice thermal crossover.
Furthermore, thermal preparation improves on the pure-ground-state approximation over much of the sampled range, although that reference remains competitive for selected low-temperature observables on the square lattices.}

\revision{As shown in Fig.~\ref{fig:shared-correlations-4x4}, the intermediate-temperature correlation discrepancies are particularly notable on the $4\times4$ square, compared with the more faithful $20\times1$ profiles.
  A plausible limitation is the shallow, one-dimensional pattern of system entangling gates, which may not readily reproduce the distribution of spatial fluctuations across both lattice directions near the thermal crossover.
Suboptimal optimization may also contribute, so these finite-size observations do not establish a unique cause or a general obstruction to two-dimensional thermal-state preparation.}

\revision{One of the main challenges in variational algorithms is the potential presence of barren plateaus in the optimization landscape, which can hinder convergence to the global optimum.
  For the two tested fixed-layer HEAs, the component-averaged gradient variances in Appendix~\ref{app:gradients} show no strong additional suppression over $N=10$--30 beyond the explicit $N^{-2}$ factor induced by objective normalization.
  Nevertheless, this finite-size evidence neither proves asymptotic trainability nor rules out barren plateaus at larger sizes~\cite{mccleanBarrenPlateausQuantum2018}.
  Improved optimizers~\cite{stokesQuantumNaturalGradient2020,kublerAdaptiveOptimizerMeasurementFrugal2020}, layer-wise training~\cite{skolikLayerwiseLearningQuantum2021}, and informed initialization~\cite{grantInitializationStrategyAddressing2019,parkHardwareefficientAnsatzBarren2024} are possible directions for further improvements, but they cannot remove a circuit's rank restriction.
Increasing depth or ancilla number instead enlarges the available circuit resources, at the expense of classical computation and additional gates on hardware.}

One potential avenue for low-temperature Gibbs state preparation is to explore problem-informed ansatzes that do not start with a maximally entangled initial state across the system--ancilla partition like the TFDA, but still leverage some prior knowledge about the system, such as its symmetries or low-energy subspace structure.
For example, one can consider initializing the ansatz with a low-energy state obtained from classical methods such as density-matrix renormalization group (DMRG)~\cite{whiteDensityMatrixFormulation1992,schollwockDensitymatrixRenormalizationGroup2011} or neural quantum states~\cite{carleoSolvingQuantumManybody2017,langeArchitecturesApplicationsReview2024}, and then applying a series of parameterized unitaries coupling the system to ancillas to introduce effective thermal fluctuations.
\revision{The starting state must be preparable at reasonable cost, and the circuit must still have sufficient entropy capacity while remaining classically tractable within the MPS framework.
The trade-offs between these factors and the resulting observable accuracy would need to be assessed.}
Another promising direction is to explore ansatzes based on other tensor network structures, such as the product spectrum ansatz~\cite{martynProductSpectrumAnsatz2019} or the multiscale entanglement renormalization ansatz (MERA)-inspired ansatz~\cite{sewellThermalMultiscaleEntanglement2022}.
\revision{For two-dimensional targets, circuits whose entangling structure follows the physical lattice more closely also merit investigation, although their longer-range gates in the MPS ordering can increase simulation costs and must be assessed separately.
}

Finally, we note that while our current work focuses on constructing the MPS representation of the purified Gibbs state by simulating the corresponding quantum circuit classically, one can also consider using tensor network tomography techniques to reconstruct the prepared state directly from measurement data obtained from quantum hardware.
\revision{MPS tomography~\cite{cramerEfficientQuantumState2010,baumgratzScalableReconstructionDensity2013,wangScalableQuantumTomography2020,gomezReconstructingQuantumStates2022,tengLearningTopologicalStates2025} provides an efficient way to learn the MPS representation of a quantum state from a limited number of measurements.
  It could therefore replace circuit simulation under suitable structural assumptions, but would require reconstruction accurate enough to evaluate the objective and its gradient for optimization.
  Its measurement requirements and reconstruction errors would therefore need to be assessed.
  In two dimensions, projected entangled-pair states (PEPS) offer a network geometry better suited to the lattice, but contractions are more costly and generally approximate, and the global von Neumann entropy is not available from a single Schmidt spectrum.
These alternatives broaden the tensor-network-assisted framework, potentially overcoming some of the limitations of MPS-based simulation.}

\revision{Overall, the targeted benchmarks studied in this work shed light on where MPS assistance is useful and where its limitations arise in variational Gibbs-state preparation.
  A classically optimized thermal preparation circuit could also supply an initial state for subsequent quantum dynamics, for example following a Hamiltonian quench, whose evolution may become difficult to simulate classically.
  This possible application lies beyond the present benchmarks and would require assessing whether the preparation accuracy is sufficient for the dynamical quantities of interest, which is left for future work.
Extending MPS-assisted Gibbs-state preparation to broader Hamiltonian classes and larger systems requires circuit capacity appropriate to the target thermal state, tractable energy and entropy evaluation, together with reliable optimization.}

\section*{\revheading{Code Availability}}

\revision{The code used to generate the results in this work is publicly available at \url{https://github.com/ruihao-li/mps-quantum-gibbs}.}

\begin{acknowledgments}
  We would like to thank Patrick Rall and Pawel Wocjan for their valuable discussions on fault-tolerant thermal state preparation.
  We are also grateful to Anirban Chowdhury and Omer Shehab for directing us to relevant papers on complexity theory.
  \revision{Finally, we thank Roland Farrell for bringing the interleaved ansatz to our attention and for helpful discussions.}
  S.V. was supported by the U.S. Department of Energy, Nuclear Physics Quantum Horizons program through the Early Career Award DE-SC0021892 and by National Science and Technology Council (NSTC) of Taiwan through Grant No. 114-2811-M-007-053.
\end{acknowledgments}

\clearpage
\bibliography{gibbs_state}

\clearpage
\onecolumngrid
\appendix

\section{\revheading{Optimization protocol and robustness}}
\label{app:optimization}

{\revcolor
  The comparisons in Secs.~\ref{sec:small-comparison} and~\ref{sec:shared-benchmarks} depend both on the states accessible to a circuit and on how reliably optimization finds them.
  In this appendix, we describe the optimization protocol used in our studies.
  We describe the common Adam--L-BFGS-B workflow, specify the differences between the small-system and shared HEA benchmarks, and then examine the variation across cold starts.
}

\subsection{\revheading{Initialization, optimization, and state selection}}
\label{app:optimization-protocol}

{\revcolor
  Both benchmarks minimize the dimensionless objective per spin $C_\beta=(\beta E-S)/N$ of Eq.~\eqref{eq:Cbeta}.
  Objective gradients are evaluated by automatic differentiation.
  Adam provides the initial optimization, followed by L-BFGS-B refinement.
  Each cold start begins from a randomly drawn parameter vector.

  For the contiguous and interleaved HEA comparisons, every angle is constrained to $[-2\pi,2\pi]$ and the Adam learning rate is set to $0.02$.
  After at least 200 Adam steps, we check progress every 25 steps by comparing the cumulative best objective with its value 50 steps earlier.
  Here the cumulative best is the lowest value encountered since initialization, rather than an average over a moving window.
  Optimization passes to L-BFGS-B after the drop in the cumulative best falls below a threshold of at most $10^{-4}$, or when the maximum number of 2000 Adam steps is reached.

  L-BFGS-B uses SciPy's native stopping tolerances, $\mathtt{ftol}=2.22\times10^{-9}$ for the scaled relative objective reduction and $\mathtt{gtol}=5\times10^{-7}$ for the infinity norm of the projected gradient of $C_\beta$.
  Either condition can end the optimization.
  We allow at most 20 line-search steps per iteration and 1500 iterations.
  States from runs that reach an iteration or evaluation limit are retained after independent reconstruction, but are distinguished from converged runs in the stopping diagnostics.
  To avoid repeated compilation, each worker reuses a JAX-compiled objective-and-gradient function across the 20 starts at a fixed circuit and inverse temperature.
  A different $\beta$ requires a separately compiled function in this implementation.

  The refined small-system TFDA--HEA comparison uses the same Adam threshold but requires two consecutive checks meeting it.
  Its 5000-step safety limit reports a failure rather than passing an unconverged run onward, and L-BFGS-B nonconvergence is likewise treated as a failure.

  For each run, we independently reconstruct both the final optimized state and the saved state with the lowest optimizer-recorded objective.
  The reported quantities use the latter state, with its energy, entropy, and $C_\beta$ recomputed independently using Qiskit and Quimb.
  For each circuit and temperature, we then select the lowest recomputed $C_\beta$ among the 20 starts.
  All observables for that circuit and temperature come from this same selected state.

  For each system, ansatz architecture, $N_a$, and $L$ in the shared HEA benchmarks, a fixed random seed generates 20 independent initial vectors with angles uniform on $[0,2\pi)$.
  The vectors are reused at every temperature for that circuit shape.
  For the small-system TFDA--HEA comparison, corresponding parent seeds generate the 20 starts for each ansatz at a given model, size, and temperature, but the ansatz-specific streams and parameter dimensions give different vectors.
  These small-system calculations also use exact dense-statevector propagation rather than MPS propagation.
}

\subsection{\revheading{Variability and convergence across cold starts}}
\label{app:optimization-diagnostics}

\begin{figure*}[t]
  \centering
  \figurefile[\textwidth]{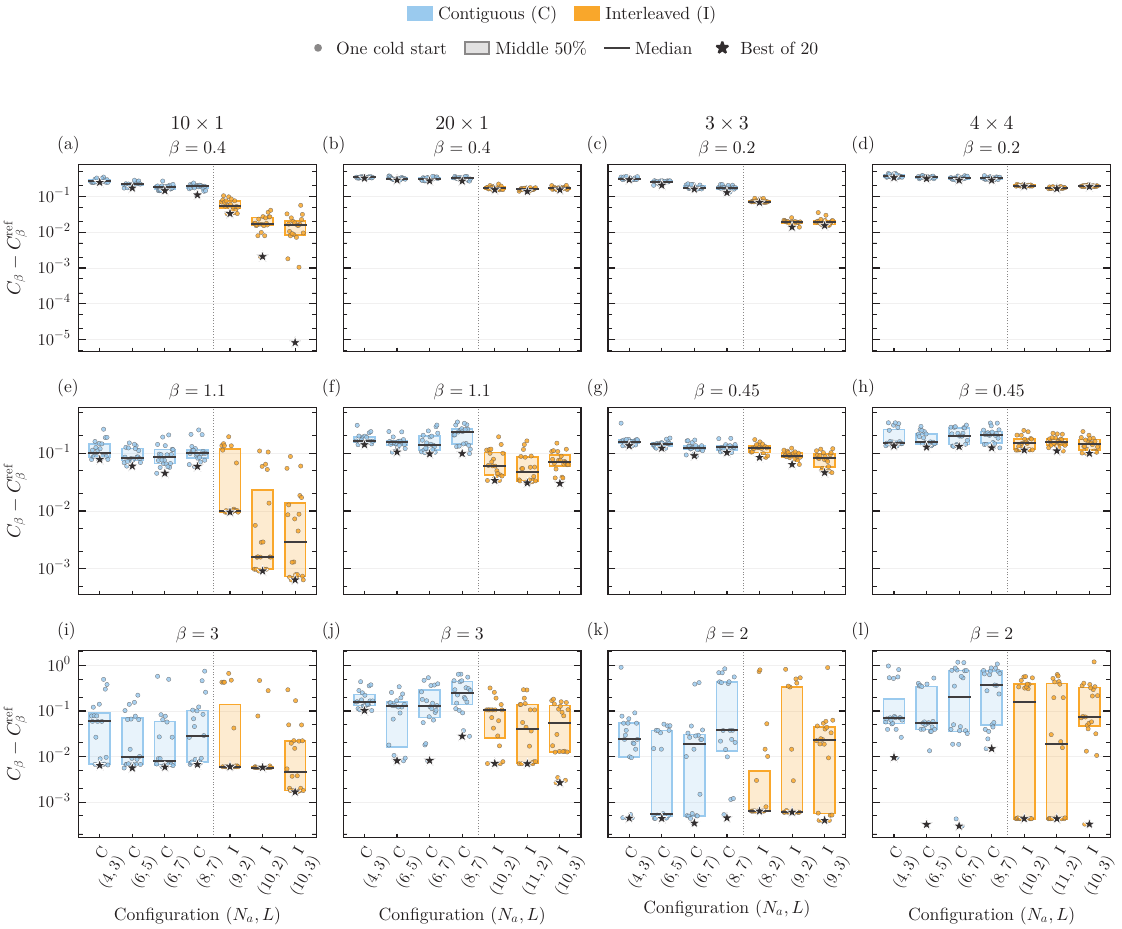}
  \caption{\revision{Run-to-run variation of the objective error $C_\beta-C_\beta^{\rm ref}$ for the open TFIM with $J=1$ and $h=0.5$.
      Here $C_\beta=(\beta E-S)/N$, with energy $E$, entropy $S$, and $N$ physical spins; $C_\beta^{\rm ref}$ is the exact thermal value.
      Columns correspond to $10\times1$, $20\times1$, $3\times3$, and $4\times4$ system; rows show representative high, intermediate, and low temperatures.
      The chain temperatures are $\beta=0.4,1.1,3$, and the square-lattice temperatures are $\beta=0.2,0.45,2$, drawn from the regimes shown in Figs.~\ref{fig:shared-correlations} and~\ref{fig:shared-correlations-4x4}.
      The inverse temperature is indicated above each panel.
      Each category denotes a contiguous (C, blue) or interleaved (I, orange) configuration with ancilla and layer counts $(N_a,L)$.
      Small dots show the independently recomputed objective errors from all 20 cold starts; horizontal offsets separate overlapping points for visibility.
  Boxes span the 25th--75th percentiles, horizontal bars mark medians, and black stars highlight the lowest recomputed objective among the 20 starts.}}
  \label{fig:optimizer-distributions}
\end{figure*}

\begin{figure*}[t]
  \centering
  \figurefile[\textwidth]{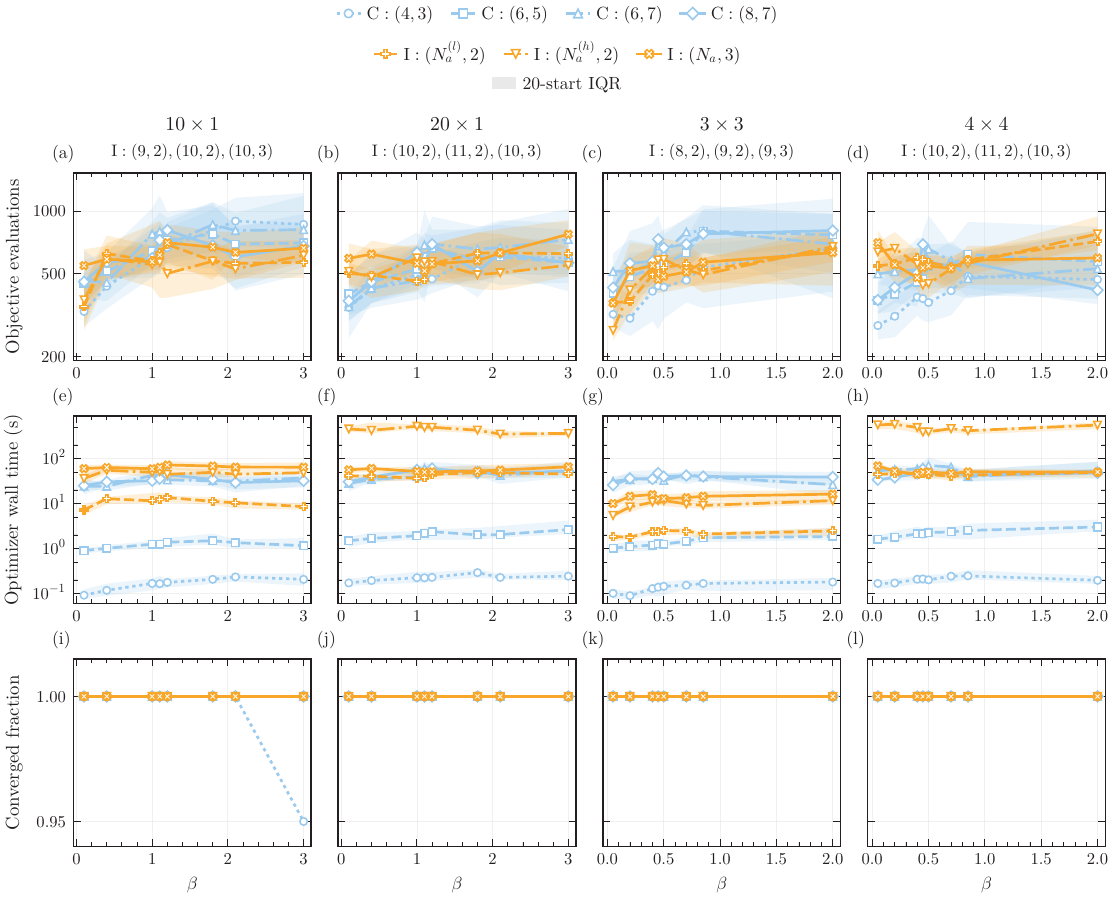}
  \caption{\revision{Optimization effort and convergence status for the open-boundary TFIM with $J=1$ and $h=0.5$, using Adam followed by L-BFGS-B.
      Columns correspond to $10\times1$, $20\times1$, $3\times3$, and $4\times4$ systems.
      From top to bottom, the rows show the total number of objective-and-gradient evaluations in panels (a)--(d), optimization wall time in seconds in (e)--(h), and the fraction of runs satisfying the convergence criteria in (i)--(l).
      Blue and orange curves denote contiguous (C) and interleaved (I) HEAs; legend pairs give ancilla and layer counts $(N_a,L)$, with $l/h$ for lower/higher interleaved ancilla budgets.
      Each circuit is optimized from 20 independent starts at each of eight inverse temperatures $\beta$.
      Evaluation and time curves show medians over the 20 starts at each temperature, with bands spanning the 25th--75th percentiles.
      Timings exclude compilation, the initial objective-only evaluation, and independent state reconstruction.
  A run is counted as converged only if it satisfies both the Adam handoff and L-BFGS-B stopping criteria.}}
  \label{fig:optimizer-effort}
\end{figure*}

{\revcolor
  The figures in the main text show the state with the lowest independently recomputed objective among the 20 starts.
  Fig.~\ref{fig:optimizer-distributions} shows how representative that selected result is by displaying all 20 objective errors for each configuration at three representative temperatures for the 1D chains and the 2D squares.
  These three selected temperatures represent the high, intermediate, and low-temperature regimes discussed in Sec.~\ref{sec:shared-benchmarks}.
  The boxes and median bars summarize typical outcomes and the stars identify the selected states.

  The separation between the best and typical outcomes varies with temperature and circuit configuration.
  In the high-temperature row, the larger systems show relatively tightly clustered objective errors, whereas the $10\times1$ interleaved circuits can have selected results substantially below their medians.
  At low temperature, several configurations show much broader distributions, particularly on the square lattices, where small selected errors coexist with much larger errors from other starts.
  Thus the accuracy of the selected state need not be typical of a single cold start.
  Conversely, tightly clustered outcomes indicate repeatability but do not establish global optimality or distinguish limited entropy capacity of the circuit from optimization that consistently reaches similar states.

  Fig.~\ref{fig:optimizer-effort} reports evaluation counts in panels (a)--(d), optimization times in (e)--(h), and stopping outcomes in (i)--(l) for the full set of 4480 optimizations, comprising four systems and seven ansatz configurations at eight temperatures with 20 starts.
  Evaluation counts add the objective-and-gradient calls made by Adam and L-BFGS-B, including line-search evaluations.
  The shaded interquartile ranges (IQRs) span the middle 50\% of the 20 runs at each temperature.
  The median evaluation counts are typically several hundred and generally increase on cooling, although the temperature dependence is not monotonic and neither architecture consistently requires fewer evaluations.
  The wall times vary much more strongly across HEA configurations, spanning several orders of magnitude: the contiguous $L=3,5$ circuits are substantially faster than the $L=7$ circuits, while the interleaved $(11,2)$ circuits take the longest on $20\times1$ and $4\times4$.
  In particular, those $(11,2)$ circuits require evaluation counts comparable to the corresponding $(10,2)$ circuits despite their much longer wall times.
  Comparable evaluation counts can therefore conceal large differences in the cost of each objective-and-gradient evaluation, consistent with the dependence on MPS bond dimensions and entropy evaluation discussed in Sec.~\ref{sec:mps-cost} and Appendix~\ref{app:mps-resources}.
  A run is counted as converged only when Adam satisfies its handoff criterion and L-BFGS-B terminates through its objective-reduction or projected-gradient tolerance, i.e., when either $\mathtt{ftol}$ or $\mathtt{gtol}$ is satisfied.
  This occurs in 4479 of the 4480 runs.
  The remaining run, with the contiguous $(4,3)$ circuit on the $10\times1$ chain at $\beta=3$, reaches the Adam's maximum iterations without meeting the handoff criterion, then satisfies the L-BFGS-B objective-reduction criterion.
  Satisfying these numerical criteria indicates termination under the specified tolerances, but it guarantees neither a local minimum nor thermal accuracy.
}

\clearpage
\section{\revheading{Exact thermal and ground-state references}}
\label{app:references}

{\revcolor
  In this appendix, we describe how the thermal references and the pure-ground-state comparator displayed in Sec.~\ref{sec:shared-benchmarks} are obtained, then use the absolute- and relative-error plots to complement the observable comparisons in Fig.~\ref{fig:shared-observables}.
}

\subsection{\revheading{Thermal and ground-state references}}
\label{app:reference-construction}

{\revcolor
  Thermal references for all four shared systems are computed exactly without a thermal-state approximation, up to floating-point numerical accuracy.
  We use a free-fermion covariance-matrix solution for 1D open-chain thermodynamics and correlations, complete dense diagonalization for $3\times3$, and complete symmetry-resolved diagonalization for $4\times4$, as described below.
  The fluctuation estimators $\chi=\beta\operatorname{Var}(M)/N$ and $c_v=\beta^2\operatorname{Var}(H)/N$ are those of Eqs.~\eqref{eq:chi} and~\eqref{eq:cv}, with total longitudinal magnetization $M=\sum_i Z_i$.

  The pure-ground-state reference is the unique finite-size even-parity ground state discussed in the main text.
  Writing its energy as $E_0$ and denoting expectation-value variance in this state by $\operatorname{Var}_0$, the comparator has
  \begin{align}
    C_\beta^{\rm GS}&=\frac{\beta E_0}{N},\\
    S_{\rm GS}&=0,\\
    \chi_{\rm GS}&=\frac{\beta}{N}\operatorname{Var}_0(M),\\
    c_{v,{\rm GS}}&=0.
  \end{align}
  This ground-state reference is computed exactly by the same methods as the thermal references, rather than by variational optimization.

  Complete diagonalization of the open $4\times4$ TFIM becomes practical by exploiting three commuting symmetries: horizontal reflection, vertical reflection, and global spin inversion.
  We classify basis states as even or odd under each of the two reflections and under global spin inversion.
  These three symmetry labels define eight independent sectors that the Hamiltonian does not mix.
  Including all sectors retains the full Hilbert space, so the symmetry decomposition reduces the computational cost without discarding any states.
  We diagonalize every real symmetric block, using all eigenvectors to compute observable expectation values.

  The eigenvalues determine the partition function $Z_\beta$ and thermal energy $E_{\rm ref}$.
  They also give the reference entropy $S_{\rm ref}=\beta E_{\rm ref}+\ln Z_\beta$, objective $C_\beta^{\rm ref}=-\ln Z_\beta/N$, and specific heat at each temperature.
  We also save the eigenstate expectation values of $M^2$ and $Z_iZ_j$, so their thermal averages give the same equal-time $\chi$ and connected correlations as in the main text.
  Global spin-inversion symmetry gives $\langle Z_i\rangle=0$.
  The connected pair correlations are averaged over physical separations with the same pair counts as in Eqs.~\eqref{eq:chain-averaged-correlation} and~\eqref{eq:square-averaged-correlation}, with no on-site terms included.
}

\begin{figure*}[t]
  \centering
  \figurefile[\textwidth]{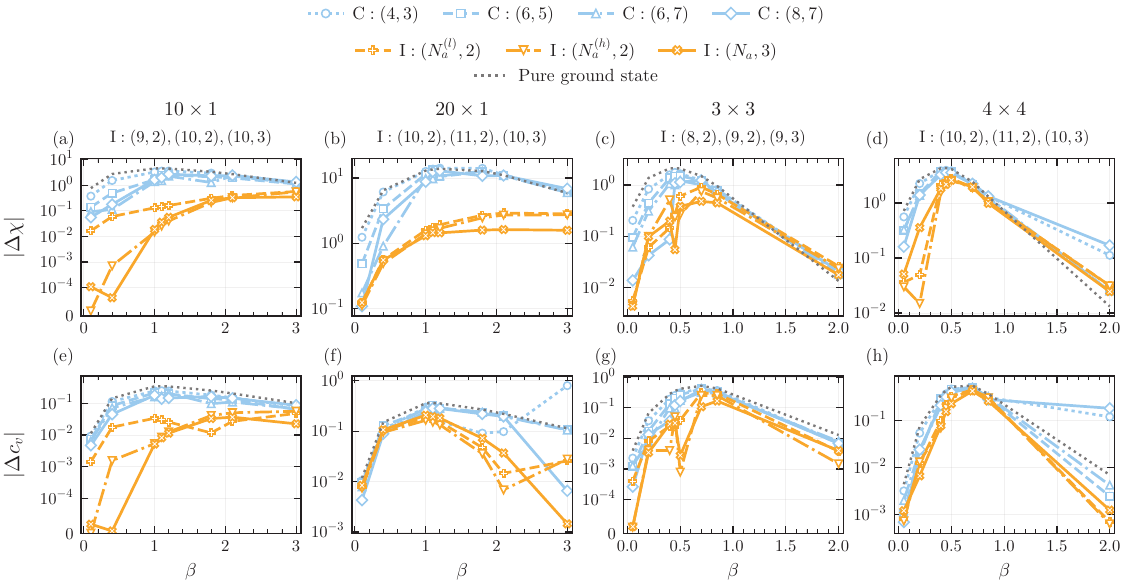}
  \caption{\normalcolor\revision{Absolute errors $|\Delta\chi|=|\chi-\chi_{\rm ref}|$ [panels (a)--(d)] and $|\Delta c_v|=|c_v-c_{v,\rm ref}|$ [panels (e)--(h)] against exact thermal references for the open-boundary TFIM with $J=1$ and $h=0.5$.
      The estimators are $\chi=\beta\operatorname{Var}(\sum_i Z_i)/N$ and $c_v=\beta^2\operatorname{Var}(H)/N$.
      Columns correspond to $10\times1$, $20\times1$, $3\times3$, and $4\times4$.
      Blue and orange curves denote contiguous (C) and interleaved (I) HEAs with ancilla and layer counts $(N_a,L)$; $l/h$ denote lower/higher ancilla budgets, with interleaved configurations above each column.
      Gray dotted curves show the exact pure-ground-state comparator.
  Each variational state is selected from 20 independent starts by the lowest independently recomputed $C_\beta$.}}
  \label{fig:ground-state-errors}

  \vspace{26pt}
  \figurefile[\textwidth]{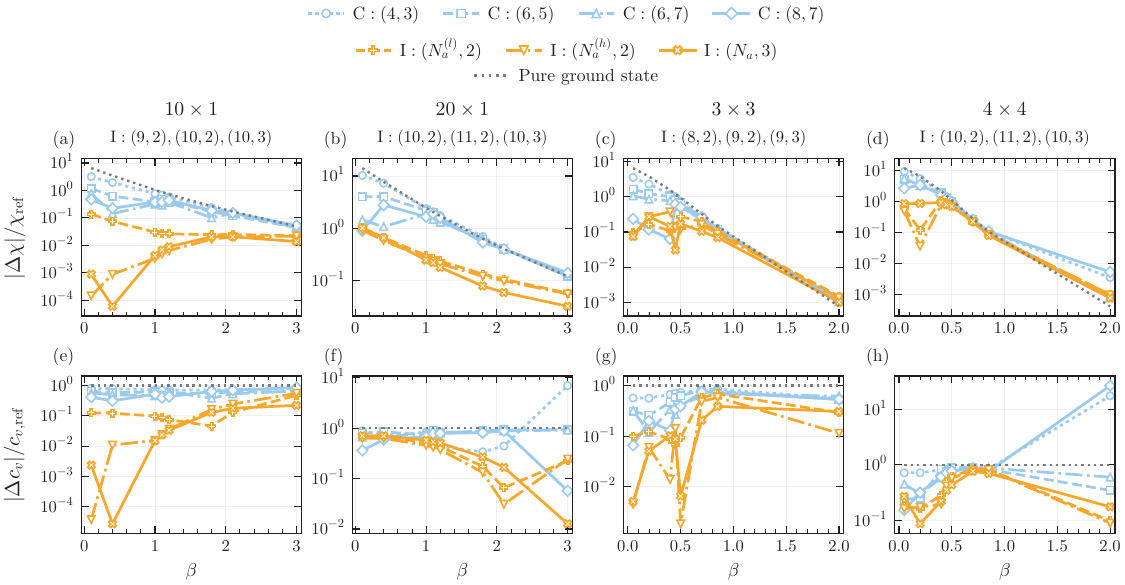}
  \caption{\normalcolor\revision{Relative errors $|\Delta\chi|/\chi_{\rm ref}$ [panels (a)--(d)] and $|\Delta c_v|/c_{v,{\rm ref}}$ [panels (e)--(h)] for the same states, systems, and legend conventions as Fig.~\ref{fig:ground-state-errors}.
      Values are fractions and all reference values are positive, with no denominator floor.
  Both figures use logarithmic vertical axes above $10^{-4}$ and a linear scale below it.}}
  \label{fig:relative-observable-errors}
\end{figure*}

\subsection{\revheading{Observable errors and ground-state comparison}}
\label{app:ground-state-observables}

{\revcolor
  Fig.~\ref{fig:ground-state-errors} resolves differences between the thermal-circuit and pure-ground-state estimates that can be difficult to distinguish in Fig.~\ref{fig:shared-observables}.
  Absolute errors show the size of the discrepancies on each observable's own scale, but curve heights alone cannot rank the fractional accuracy of two observables with different scales.
  The companion relative errors in Fig.~\ref{fig:relative-observable-errors} measure the discrepancies as fractions of the corresponding exact thermal values.
  The explicit factors of $\beta/N$ in $\chi$ and $\beta^2/N$ in $c_v$ cancel in these ratios, which therefore also give the fractional errors in the underlying magnetization and energy variances.
  At fixed finite size, the finite spectral ranges of $M$ and $H$ impose state-independent bounds on their variances.
  The explicit factors of $\beta$ and $\beta^2$ therefore force the absolute estimator errors to vanish as $\beta\to0$, without requiring accurate magnetization or energy variances.
  Conversely, a small thermal $c_v$ at low temperature can make a modest absolute discrepancy large in relative terms.
  The two figures should therefore be read together.

  Several absolute-error curves have their largest deviations at intermediate temperatures, particularly on the square lattices, but this is not a universal measure of where thermal preparation is least accurate.
  For example, the interleaved $20\times1$ circuits have their largest relative errors in $\chi$ at the highest temperature and improve steadily on cooling [Fig.~\ref{fig:relative-observable-errors}(b)], even though their absolute errors initially increase [Fig.~\ref{fig:ground-state-errors}(b)].
  For the interleaved $4\times4$ circuits, however, both $\chi$ and $c_v$ retain pronounced intermediate-temperature deviations on the relative scale [Fig.~\ref{fig:relative-observable-errors}(d) and (h)].
  These deviations are therefore not simply a consequence of multiplying variance errors by $\beta$ or $\beta^2$.
  The target entropy is largest at high temperature, but the entropy-capacity bound alone does not determine the accuracy of either variance, or the temperature at which its error is largest.

  The two observables also need not improve together on cooling.
  In particular, the interleaved $4\times4$ estimates of $\chi$ become much more accurate while $c_v$ remains substantially underestimated, before both improve at the coldest point.
  On the smaller $10\times1$ chain, the interleaved circuits reproduce the thermal $c_v$ peak well, but their fractional errors are larger at the coldest point [Fig.~\ref{fig:relative-observable-errors}(e)].
  At that temperature, the exact $c_v$ value is much smaller, so the absolute deviations should also be consulted when assessing their significance [Fig.~\ref{fig:ground-state-errors}(e)].
  Therefore, these are complementary tests of the prepared state's magnetization and energy distributions, but not evidence that one observable is uniformly harder to reproduce than the other.

  Let us now compare the variational estimates with the pure-ground-state reference.
  Across the four systems, two observables, and eight temperatures, at least one displayed variational candidate has a smaller error than the pure-ground-state prediction in 62 of 64 comparisons.
  This comparison is unchanged on passing from absolute to relative errors, since all candidates at a given system, observable, and temperature share the same positive denominator.
  The two exceptions are $\chi$ at $\beta=2$ for $3\times3$ and $4\times4$.
  The smallest displayed variational absolute errors are $0.0175$ and $0.0245$, compared with ground-state errors $0.0133$ and $0.0132$, respectively.
  At both points, a variational $c_v$ estimate remains closer to the thermal reference.
  Since $c_{v,{\rm GS}}=0$, the ground-state relative error in $c_v$ is exactly one throughout the sampled range [Fig.~\ref{fig:relative-observable-errors}(e)--(h)].
  Its peak in the absolute-error figure simply follows the exact thermal $c_v$ rather than indicating a change in the fractional accuracy of this comparator.

  As discussed in Sec.~\ref{sec:shared-benchmarks}, at the coldest square-lattice points, the two nearly degenerate lowest states are populated almost equally, giving a thermal entropy close to $\ln2$.
  Their magnetization fluctuations are very similar, so mixing them leaves $\chi$ close to its ground-state value.
  However, their tiny energy splitting contributes very little to the energy variance.
  The nonzero thermal $c_v$ instead comes mainly from the small populations of higher excited states, whose larger energy separations make their contribution to energy fluctuations significant.
  Thermal preparation is nevertheless useful over much of the sampled range, but not every variational estimate significantly improves on the pure ground state.
}

\clearpage
\section{\revheading{MPS resources and numerical checks}}
\label{app:mps-resources}

{\revcolor
  Sec.~\ref{sec:mps-cost} compares the general time and memory requirements of the two layouts.
  Fig.~\ref{fig:mps-resources} complements that analysis with the achieved bond dimensions and computation times for the four-system benchmarks.
  Here we explain the bond caps, the independent reconstruction used to check the optimized states, and the numerical regularization of entropy.
}

\subsection{\revheading{Bond dimensions and truncation checks}}
\label{app:bond-dimensions}

\begin{figure*}[t]
  \centering
  \figurefile[\textwidth]{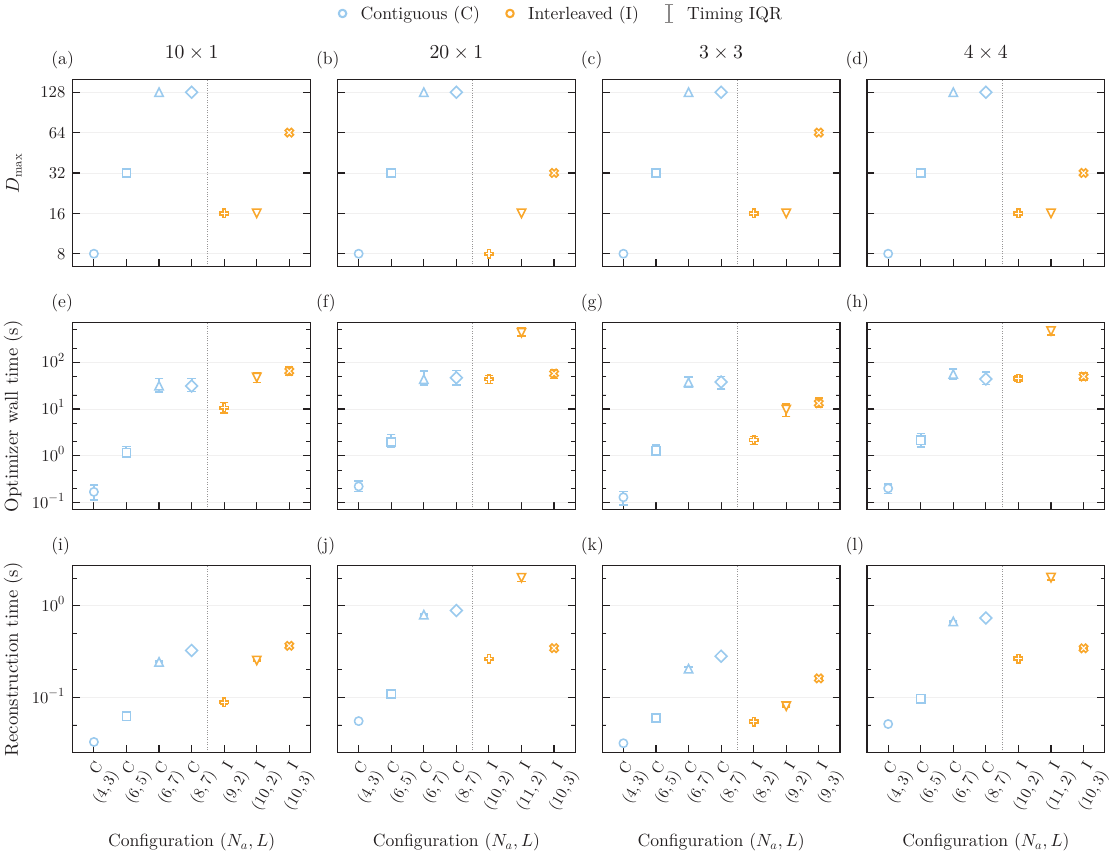}
  \caption{\revision{MPS bond dimensions and computation times for the open-boundary TFIM with $J=1$ and $h=0.5$.
      Columns correspond to $10\times1$, $20\times1$, $3\times3$, and $4\times4$ systems.
      Rows show (a)--(d) the largest MPS bond dimension $D_{\max}$ across the bonds of each selected state, (e)--(h) optimization wall time in seconds, and (i)--(l) independent reconstruction wall time in seconds, with a common logarithmic scale within each row.
      Blue and orange markers denote contiguous (C) and interleaved (I) HEAs, with ancilla and layer counts $(N_a,L)$ on the horizontal axes.
      For each configuration, the achieved bond dimension is the same at all eight temperature points, with each state selected from 20 starts by the lowest recomputed $C_\beta$.
      Timing markers show medians over all 160 starts per configuration, with error bars spanning the 25th--75th percentiles (interquartile ranges).
  The calculations allow bond dimensions up to $2^L$ for contiguous and $4^L$ for interleaved circuits, with zero truncation cutoff.}}
  \label{fig:mps-resources}
\end{figure*}

{\revcolor
  To distinguish limitations of the ansatz from truncation in the MPS calculation, we use zero truncation cutoff and maximum bond dimensions sufficient for the circuit: $2^L$ for the contiguous HEA and $4^L$ for the interleaved HEA\@.
  In the four-system benchmarks in Sec.~\ref{sec:shared-benchmarks} of the main text, the contiguous configurations $(N_a,L)=(4,3),(6,5),(6,7),(8,7)$ use bond caps $D=8,32,128,128$, respectively.
  All interleaved configurations use $D=16$ for $L=2$ and $D=64$ for $L=3$, irrespective of the ancilla number within the tested configurations.
  These configured caps are sufficient bounds, not necessarily the bond dimensions attained by individual states.
  Fig.~\ref{fig:mps-resources}(a)--(d) reports the achieved maximum bond dimensions at the selected optimized states, with optimization and reconstruction times in panels (e)--(h) and (i)--(l), respectively.
  For example, the selected interleaved $(10,2)$ states on $20\times1$ have maximum bond dimension eight, below their configured cap of 16.
  The bond dimension needed along the MPS ordering should also not be identified with the rank of the entire system's reduced density matrix: interleaved registers can have large system--ancilla entanglement distributed across several bonds, as explained in Sec.~\ref{sec:mps-cost}.

  The independently reconstructed states report zero discarded weight, consistent with the sufficient circuit bounds.
  The configured bond dimension caps therefore introduce no additional truncation beyond the circuit's own restrictions, up to floating-point numerical accuracy.
  Raising them further would not enlarge the set of states accessible to the same circuit.
}

\subsection{\revheading{Independent reconstruction and computation times}}
\label{app:timing}

{\revcolor
  Independent reconstruction checks the thermodynamic quantities without reusing the differentiable MPS calculation employed during optimization.
  We build the same circuit in Qiskit, bind the saved parameters, and propagate it as an MPS using Quimb to recompute $E$, $S$, and $C_\beta$.
  This is a therefore classical reevaluation at fixed parameters.
  Both the final state and the saved state with the lowest optimizer-recorded objective are reconstructed for each run.
  As described in Appendix~\ref{app:optimization}, the latter supplies the reported run result, and the lowest recomputed $C_\beta$ among the 20 starts selects the state used for the observable comparisons.

  Fig.~\ref{fig:mps-resources}(e)--(l) summarizes the optimization and reconstruction wall times over 160 starts per configuration, comprising 20 starts at each of eight temperatures.
  Markers show medians, and error bars span the 25th--75th percentiles.
  Optimization times exclude compilation, the initial objective-only evaluation, independent reconstruction, and writing the results; reconstruction times include propagation and thermodynamic reevaluation of both saved states from each run.
  The runs used a 12-core arm64 machine with 24 GiB of memory.
  Fig.~\ref{fig:optimizer-effort}(e)--(h) resolves the optimization times by temperature.

  Within the contiguous family, the deeper $L=7$ circuits have much larger bond dimensions and longer computation times than the $L=3,5$ circuits.
  For interleaved circuits, the smaller bond dimensions do not necessarily imply shorter optimization times: the $(11,2)$ configurations on $20\times1$ and $4\times4$ take the longest among the displayed candidates.
  On $4\times4$, increasing $N_a$ from 10 to 11 at $L=2$ leaves the achieved maximum bond dimension unchanged, yet both median times increase substantially.
  This distinction is consistent with the additional ancilla-dependent entropy cost discussed in Sec.~\ref{sec:mps-cost}.
  Median reconstruction times are much shorter than optimization times for every displayed configuration, since reconstruction reevaluates two fixed states rather than repeatedly evaluating the objective and its gradient while varying the parameters.
  Its interquartile ranges are included in panels (i)--(l), but are too small to visualizef clearly on the logarithmic scale.
}

\subsection{\revheading{Entropy regularization}}
\label{app:entropy-regularization}

{\revcolor
  Entropy evaluation requires particular care when the reduced density matrix has very small eigenvalues.
  An eigenvalue's contribution $-p\ln p$ vanishes as $p\to0^+$, whereas its derivative $-(\ln p+1)$ diverges.
  Small spectral weights can therefore have little effect on the entropy value while still making its gradient sensitive to numerical errors.
  Roundoff can also produce small negative eigenvalues, for which the real logarithm is undefined.
  We use float64 arithmetic and explicit regularization to handle these issues.

  For contiguous circuits, entropy is obtained from Gram matrices of overlaps within the two MPS blocks at the system--ancilla boundary.
  A $10^{-14}$ diagonal shift stabilizes one Gram matrix.
  Negative spectral weights are clipped to zero, the weights are normalized, and weights at or below $10^{-15}$ are omitted from the entropy sum, with a guard against $\ln0$.
  For interleaved circuits and the small exact-statevector TFDA--HEA benchmark, eigenvalues at or below $\epsilon=10^{-14}$ are discarded and the retained spectrum is renormalized before evaluating entropy.
  The gradient calculation includes the dependence of this normalization on the circuit parameters, so it differentiates the regularized expression rather than applying the singular unregularized derivative to discarded modes.

  These spectral operations do not truncate the propagated MPS.
  They leave the circuit family unchanged but modify the numerical objective and its gradients, so they can influence the optimization.
  However, the regularization is small enough that it does not significantly affect the reported results.
}

\clearpage
\section{\revheading{Controlled finite-size gradient-variance analysis}}
\label{app:gradients}

\begin{figure*}[t]
  \centering
  \figurefile[0.680\textwidth]{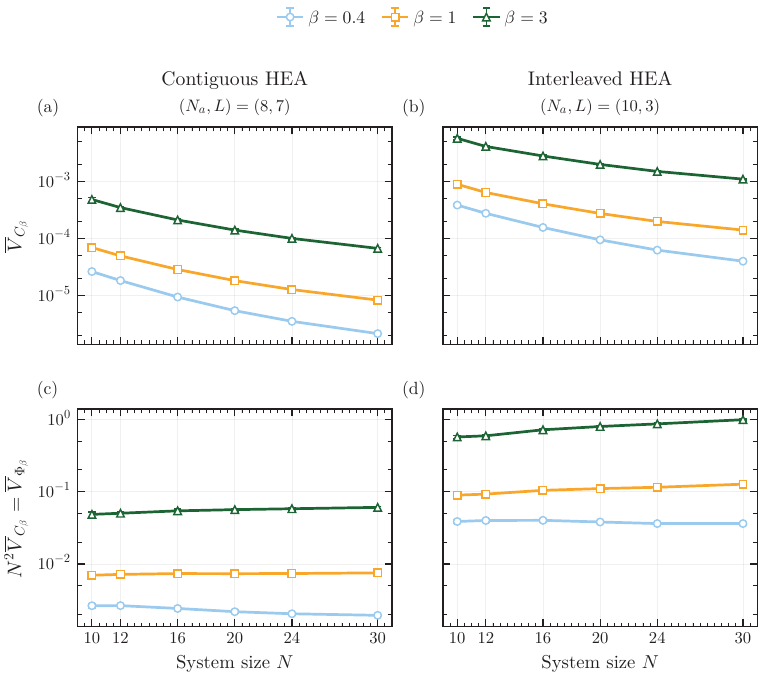}
  \caption{\revision{System-size dependence of objective-gradient variances for the open TFIM chain with $J=1$ and $h=0.5$.
      Panels (a) and (c) use the contiguous HEA with $(N_a,L)=(8,7)$, and panels (b) and (d) use the interleaved HEA with $(10,3)$, where $N_a$ and $L$ are the ancilla and layer counts.
      The upper panels show the average variances $\overline V_{C_\beta}$ of the normalized objective $C_\beta$.
      Unbiased sample variances are evaluated over 128 independent parameter vectors drawn uniformly from $[0,2\pi)$ for each circuit size, without optimization.
      Blue circles, orange squares, and green triangles correspond to $\beta=0.4,1,3$, using the same vectors at all three temperatures.
      The lower panels show $\overline V_{\Phi_\beta}$, where $\Phi_\beta$ removes the explicit normalization factor.
      Error bars give pointwise 95\% percentile intervals from 5000 bootstrap resamples of whole paired energy--entropy gradient vectors.
  Connecting lines are visual guides.}}
  \label{fig:gradient-scaling}
\end{figure*}

{\revcolor
  Appendix~\ref{app:optimization} examines how the optimized objective values vary across 20 independent random initializations.
  These outcomes show how consistently optimization finds low-cost states, but do not establish whether a barren plateau is present in the optimization landscape, that is, whether the objective gradients become exponentially suppressed as the system size increases.
  Therefore, in this appendix, we examine whether gradients of the actual optimization objective, $C_\beta=(\beta E-S)/N$, show additional exponential suppression with system size beyond its explicit normalization over the accessible range $N=10$--30.
  We use the same open one-dimensional TFIM with $J=1$ and $h=0.5$ as in the HEA benchmarks, but we do not perform any optimization.
  For each circuit shape, we evaluate gradients at 128 independent random parameter vectors.
  The sizes are $N\in\{10,12,16,20,24,30\}$ and the inverse temperatures are $\beta\in\{0.4,1,3\}$.
  We choose the deepest tested representative of each HEA family: contiguous $(N_a,L)=(8,7)$ and interleaved $(10,3)$.
  Their resources are not identical, so we examine the size dependence within each family separately.

  Each circuit has $P=(N+N_a)L$ parameters.
  For each architecture and size, NumPy's PCG64 generator draws vectors uniformly from $[0,2\pi)^P$ using a fixed seed.
  We save the energy and entropy gradients separately and form $\nabla C_\beta=(\beta\nabla E-\nabla S)/N$ for each temperature.
  The same random states are used at all three temperatures.
  Writing $\partial_\mu=\partial/\partial\theta_\mu$, we report the variance across parameter vectors, averaged over components:
  \begin{equation}
    \overline V_{C_\beta}(N)=\frac1P\sum_{\mu=1}^{P}
    \operatorname{Var}_{\boldsymbol\theta}
    [\partial_\mu C_\beta],
    \label{eq:gradient-variance}
  \end{equation}
  where $\operatorname{Var}_{\boldsymbol\theta}$ refers to variation over the random parameter samples and is estimated using the unbiased sample variance.
  To separate the explicit size normalization from the behavior of the circuit gradients, we also define the extensive objective $\Phi_\beta=\beta E-S$.
  Its component-averaged gradient variance, $\overline V_{\Phi_\beta}$, obeys the exact identity
  \begin{equation}
    \overline V_{\Phi_\beta} = N^2\overline V_{C_\beta}.
    \label{eq:gradient-normalization}
  \end{equation}
  Fig.~\ref{fig:gradient-scaling}(a) and (b) show that the gradient variance of the normalized objective decreases with $N$ for both HEAs at all three temperatures.
  However, Eq.~\eqref{eq:gradient-normalization} shows that dividing the objective by $N$ introduces an exact $N^{-2}$ factor into its gradient variance; this factor alone is therefore not evidence of a barren plateau.
  Once it is removed, the variances in panels (c) and (d) change much less with system size.
  At $\beta=0.4$ they decrease modestly between $N=10$ and 30, whereas at $\beta=1,3$ they remain comparable or increase.

  The endpoint ratio $\overline V_{\Phi_\beta}(30)/\overline V_{\Phi_\beta}(10)$ also summarizes this contrast.
  At $\beta=0.4,1,3$, respectively, it is $0.736$, $1.079$, and $1.251$ for the contiguous HEA and $0.935$, $1.417$, and $1.735$ for the interleaved HEA\@.
  Thus the substantial decrease in the upper row is largely explained by the intensive normalization, rather than a comparable suppression of the extensive-objective gradients.

  The figure's error bars are pointwise 95\% percentile intervals from 5000 bootstrap resamples of the 128 whole paired energy--entropy gradient vectors at each architecture and size.
  Resampling whole vectors preserves correlations among parameter components and between energy and entropy gradients.
  The intervals are narrow on the logarithmic scale, with relative half-widths of approximately 3--9\% of the estimated variances.
  This indicates modest sampling uncertainty in the component-averaged variances and supports the broad contrast between the decreasing normalized variances and the much weaker size dependence after removing the normalization factor.

  Taken together, Fig.~\ref{fig:gradient-scaling} shows no strong additional suppression of this component-averaged variance beyond the explicit normalization over $N=10$--30 for either tested HEA.
  However, this finite-size evidence does not establish asymptotic trainability, nor does it completely rule out barren plateaus at even larger $N$.
}

\clearpage

\end{document}